\documentclass{birkmult}
\usepackage{amsfonts,amssymb,amsmath}
\newtheorem{theorem}{Theorem}
\newtheorem{lemma}[theorem]{Lemma}
\newtheorem{proposition}[theorem]{Proposition}
\newtheorem{corollary}[theorem]{Corollary}

\theoremstyle{definition}

\newtheorem{problem}[theorem]{Problem}

\theoremstyle{remark}
\newtheorem{remark}[theorem]{Remark}

\numberwithin{equation}{section}
\numberwithin{theorem}{section}

\def\piz{\pi}         
\def\what{\widehat}

\def\ot{\otimes}

\def\d{\partial}

\def\vac{|0\rangle}       
\def\CC{\mathbb{C}}       
\def\ZZ{\mathbb{Z}}       
\def\RR{\mathbb{R}}       
\def\NN{\mathbb{N}}       

\def\la{\lambda}
\def\ph{\varphi}
\def\al{\alpha}
\def\be{\beta}

\def\sl{s\ell}
\def\Tens{{\mathcal{T}}} 
\def\Sym{{\mathcal{S}}}  
\DeclareMathOperator{\Span}{span}
\DeclareMathOperator{\Res}{Res}

\DeclareMathOperator{\End}{End}
\DeclareMathOperator{\Reg}{Reg}
\DeclareMathOperator{\Zhu}{Zhu}
\newcommand{\mc}[1]{{\mathcal #1}}
\newcommand{\mf}[1]{{\mathfrak #1}}

\newcommand{\lt}{\boldsymbol\ell.\boldsymbol{t}.\,}

\begin{document}

\title{Non-linear Lie conformal algebras with three generators}

\dedicatory{Dedicated to our teacher Victor Kac on the occasion
of his 65th birthday}

\author{Bojko Bakalov}

\thanks{The first author was partially supported by NSF grant DMS-0701011}

\address{Department of Mathematics \\
North Carolina State University \\
Raleigh, NC 27695, USA}

\email{bojko\_bakalov@ncsu.edu}

\author{Alberto De Sole}

\address{Dipartimento di Matematica \\
Universit\`a di Roma ``La Sapienza'' \\
Piaz.le Aldo Moro 5 \\
00185 Roma, Italy}

\email{desole@mat.uniroma1.it}


\begin{abstract}
We classify certain non-linear Lie conformal 
algebras with three generators, which can be viewed as deformations of the
current Lie conformal algebra of $\sl_2$.
In doing so we discover an interesting 
$1$-parameter family of non-linear Lie conformal 
algebras $R_{-1}^d$ $(d\in\NN)$
and the corresponding freely generated vertex algebras $V_{-1}^d$,
which includes for $d=1$
the affine vertex algebra of $\sl_2$ at the critical level $k=-2$.
We construct free-field realizations of the algebras $V_{-1}^d$
extending the Wakimoto realization of $\what{\sl}_2$ at the 
critical level, and we compute their Zhu algebras.
\end{abstract}

\maketitle



\section{Introduction}\label{s-intro}

\subsection{Vertex algebras and Lie conformal algebras}

The notion of a \emph{vertex algebra} \cite{B}
provides an axiomatic algebraic description of the 
operator product expansion of chiral fields in 2-dimensional 
conformal field theory. Vertex algebras played an important role
in the conceptual understanding of the ``monstrous moonshine''
\cite{CN,FLM,B2,G}, and have also proved useful
in the representation theory of infinite-dimensional Lie algebras.

The data of a vertex algebra consist of the space of \emph{states} $V$ 
(an arbitrary vector superspace), the \emph{vacuum vector} $\vac \in V$,
the infinitesimal \emph{translation operator} $T \in \End V$, and a
collection $\mc F$ of $\End V$-valued \emph{quantum fields}, 
subject to the axioms formulated below (which
are ``algebraic'' consequences of Wightman's axioms);
see \cite{FLM,K,FB,LL,DSK2}.
The quantum fields are linear maps
from $V$ to $V[[z]] [z^{-1}]$, where $z$ is a formal variable,
and can be viewed as formal series
\begin{equation*}
a(z) = \sum_{n\in\ZZ} a_{(n)} z^{-n-1} \,,
\qquad a_{(n)} \in \End V \,,
\end{equation*}
such that $a_{(n)} b = 0$ for $n\gg0$ (i.e., for $n$ large enough).

The four axioms of a vertex algebra $\bigl(V$, $\vac$, $T$, 
$\mc F = \{ a^\al(z) 
\}\bigr)$ are:
\begin{eqnarray*}
     \hbox{(vacuum axiom)}   && T \vac = 0\, , 
\\[1ex]
     \hbox{(translation covariance)} && [T,a^\al (z) ]
        = \partial _z a^\al(z) \, , 
\\[1ex]
     \hbox{(locality)} && (z-w)^{N} [a^\al(z) , a^\be (w)]=0
        \;\hbox{ for }\, N \gg 0 \, , 
\\[1ex]
     \hbox{(completeness)} && \hbox{all vectors}\;\; 
    a^{\al_1}_{(n_1)} \cdots a^{\al_s}_{(n_s)} \vac \;\; 
    \hbox {linearly span}\; V\, .
\end{eqnarray*}

If we enlarge $\mc F$ to the 
maximal collection $\bar{\mc F}$ of quantum fields for which the axioms
still hold, then the map 
\begin{equation*}
\bar{\mc F} \to V \,, \quad 
a(z) \mapsto a_{(-1)} \vac
\end{equation*}
is bijective (see e.g.\ \cite{K,DSK2}). We thus get the
\emph{state-field correspondence}, defined as the inverse map
\begin{equation*}
 V \to \bar{\mc F} \,, \quad  
a \mapsto Y (a,z) = \sum_{n\in\ZZ} a_{(n)} z^{-n-1} \,.
\end{equation*}
Here and further, we use the customary notation $Y(a,z)$ for the quantum
field corresponding to the state $a \in V$. 
The state-field correspondence allows one to introduce bilinear
products on $V$ for each $n \in \ZZ$ by letting
\begin{equation*}
  a_{(n)} b = \Res_z z^{n} \, Y (a,z)b\,,
\end{equation*}
where the formal \emph{residue} denotes the coefficient of $z^{-1}$.

The original Borcherds definition \cite{B} of a vertex algebra 
was formulated by taking a vector space $V$ with the vacuum vector $\vac$ 
and bilinear products $a_{(n)}b$ for each $n \in \ZZ$, satisfying a
simple vacuum identity and other more complicated identities, which
can be combined into one cubic identity called the 
\emph{Borcherds identity} 
(see \cite[Eq.\ (4.8.3)]{K} and also \cite{FLM,FB,LL}). 
This identity is somewhat similar to the Jacobi identity, and it is as
important for the theory of vertex algebras as the latter is for
the theory of Lie algebras.

The \emph{Wick product} (= normally ordered product) 
in a vertex algebra $V$ is defined as
${:} ab {:} = a_{(-1)} b$.
Some elementary but useful consequences of the axioms are:
\begin{equation}\label{vac2} 
a_{(n)} \vac = 0 \,, \quad {:} a \vac {:} = a \,, \quad
a_{(-n-1)}b = \frac1{n!} \, {:} (T^n a) b {:}
\,, \qquad a,b \in V \,, \; n\in\ZZ_+ \,,
\end{equation}
where $\ZZ_+$ denotes the set of non-negative integers.
An important special case of the Borcherds identity is 
the \emph{commutator formula}
\begin{equation}\label{comfla}
[Y(a,z), Y(b,w)] = \sum_{j\in\ZZ_+} Y(a_{(j)} b,w) \, 
\d_w^j \delta(z-w) / j! \,,
\end{equation}
where 
\begin{equation*}
\delta(z-w) = \sum_{m\in\ZZ} z^{-m-1} w^m
\end{equation*}
is the formal delta-function 
(note that the sum in the right-hand side of \eqref{comfla} is finite because
$a_{(j)} b = 0$ for $j\gg0$).
%
Formula \eqref{comfla} is conveniently encoded by the 
\emph{$\la$-bracket} \cite{K,DK}
\begin{equation*}
[a_\la b] 
= \Res_z e^{z\la} \, Y(a,z)b
= \sum_{j\in\ZZ_+} \frac{\la^j}{j!} \, a_{(j)} b \,,
\end{equation*}
which is a polynomial in $\la$.

In any vertex algebra, the $\la$-bracket
satisfies the axioms of a 
\emph{Lie conformal} (\emph{super})\emph{algebra} \cite{K}
(also known as a ``vertex Lie algebra'' \cite{P,FB,DLM2}).
This is a $\CC[T]$-module $R$
with a $\CC$-linear map $R\ot R \to R[\la]$, subject to the following axioms:  
\begin{eqnarray}
\label{ca}
\vphantom{\Big(}
\text{(sesqui-linearity)} &&
[(T a)_\la b] \,=\, -\la [a_\la b] \,, \quad
[a_\la (T b)] \,=\, (\la+T) [a_\la b] \,, \quad
\\
\label{ss}
\vphantom{\Big(}
\text{(skew-symmetry)} &&
[a_\la b] = -(-1)^{p(a) p(b)} \, [b_{-\la-T} a] \,, 
\\
\label{jac}
\vphantom{\Big(}
\text{(Jacobi identity)} &&
[[a_\la b]_{\la+\mu} c] = [a_\la[b_\mu c]] 
- (-1)^{p(a) p(b)} \, [b_\mu[a_\la c]] \,,
\end{eqnarray}
where $p(a)\in\ZZ/2\ZZ$ denotes the parity of an element $a$.
Simple Lie conformal superalgebras were classified 
in \cite{DK,FK}; their representation theory and cohomology theory 
were developed in \cite{CK,BKV} and other works.

The $\la$-bracket 
and Wick product in a vertex algebra are related by the following identities
\cite{K,BK}:
\begin{gather}
\intertext{(quasi-commutativity)}
\label{qcom}
{:} ab{:}  - (-1)^{p(a)p(b)} \, {:} ba{:}  
= \int_{-T}^0 \, d\la \, [a_\la b] \,,
\intertext{(quasi-associativity)}
\label{qas}
{:} ({:} ab{:} )c{:}  - {:} a({:} bc{:} ){:} 
 =  {:} \Bigl( \int_0^T d\la \, a \Bigr)[b_\la c]{:} 
 + (-1)^{p(a)p(b)} \,
   {:} \Bigl( \int_0^T d\la \, b \Bigr)[a_\la c]{:}  \,,
\intertext{(noncommutative Wick formula)} 
\label{wick}
[a_\la {:} bc{:} ] 
  = {:} [a_\la b] c{:}  + (-1)^{p(a)p(b)} \, {:} b [a_\la c]{:} 
  + \int_0^{\la} \, d\mu \, [[a_\la b]_\mu c] \,.
\end{gather}
Conversely, the above formulas, together with the axioms of a
Lie conformal algebra for $[a_\la b]$ (and the vacuum and translation
covariance properties of the Wick product), provide an equivalent
definition of the notion of a vertex algebra~\cite{BK}.

Throughout the paper, we will use the standard convention 
that a normally ordered product of more than two factors is taken 
from right to left; for example
\begin{equation}\label{mono2}
{:} abc {:} = {:} a ({:} bc {:}) {:} \,, \qquad
{:} abcd {:} = {:} a ({:} bcd {:}) {:}
= {:} a ({:} b ({:} c d {:}) {:}) {:} \,.
\end{equation}
{}From \eqref{qcom} and \eqref{qas}, one obtains the 
following useful identity \cite{BK}:
\begin{equation}\label{qcom2}
{:} abc {:} - (-1)^{p(a)p(b)} \, {:} bac {:}
= {:} \Bigl( \int_{-T}^0 \, d\la \, [a_\la b] \Bigr) c {:} \,.
\end{equation}
We are also going to need the \emph{right Wick formula} \cite{BK}:
\begin{equation}\label{rwick}
\begin{split}
[{:} ab {:}_\la c]
= {:} (e^{T \d_\la} a) [b_\la c] {:} 
&+ (-1)^{p(a)p(b)} \, {:} (e^{T \d_\la} b)[a_\la c] {:} 
\\
&+ (-1)^{p(a)p(b)} \, \int_0^\la d\mu \, [b_\mu[a_{\la-\mu} c]] \,,
\end{split}
\end{equation}
which can be derived from \eqref{ss} and \eqref{wick}.

\subsection{Non-linear Lie conformal algebras}

The relationship between Lie conformal algebras and vertex algebras
is somewhat similar to the one between Lie algebras and 
associative algebras.
In particular, to any Lie conformal algebra $R$ one canonically associates
a vertex algebra $V(R)$
known as the \emph{universal enveloping} vertex algebra of $R$
(see \cite{K,BK,GMS}).
In this way one can obtain, for instance, the vertex algebras 
associated to representations of the \emph{Virasoro} algebra or
\emph{affine Kac--Moody} algebras \cite{FZ,K,FB,LL}.
These vertex algebras have the property that they are generated
by a finite collection of fields whose $\la$-brackets are \emph{linear} 
combinations of the same fields and their derivatives. However, there are
many examples in which one has a \emph{non-linear} relationship,
i.e., the $j$-th products $a_{(j)} b$ ($j\in\ZZ_+$) of elements 
$a,b \in R$ do not necessarily belong to $R$ but are 
obtained by taking normally ordered products of elements from $R$. 
An important class of such examples is provided by \emph{$W$-algebras};
see \cite{Za,FF1,FKW,BS,FB,DSK2} and the references therein.

This has motivated the notion of a \emph{non-linear Lie conformal algebra}
\cite{DSK1} as a $\CC[T]$-module $R$ with a $\la$-bracket
$R\ot R \to \CC[\la] \ot \Tens(R)$, 
where $\Tens(R)$ is the tensor algebra over $R$, 
all tensor products being over $\CC$.
In order to be able to use induction arguments, one assumes that
$R = \bigoplus_{\Delta\in \frac12\ZZ_+} R[\Delta]$
is graded so that
\begin{equation}\label{wtla}
\Delta(Ta)=\Delta(a) + 1 \,, \qquad
\Delta(a_{(j)}b)=\Delta(a)+\Delta(b)-j-1 
\end{equation}
for all $j\in\ZZ_+$.
When $a\in R[\Delta]$ we say that $a$ has \emph{conformal weight} $\Delta$
and we use the notation $\Delta(a)=\Delta$.
The $\la$-bracket $[a_\la b]$ should then satisfy axioms 
\eqref{ca} and \eqref{ss}.
Moreover, in order to impose the 
Jacobi identity \eqref{jac} we need to extend the $\la$-bracket
to the whole tensor algebra $\Tens(R)$ (see \cite{DSK1}).
We first define the Wick product
of $a \in R$ and $B \in \Tens(R)$ as ${:} aB {:} = a \ot B$,
and we then extend the Wick product and $\la$-bracket to the whole $\Tens(R)$
using quasi-associativity \eqref{qas}
and the Wick formulas \eqref{wick} and \eqref{rwick}.
Then the Jacobi identity \eqref{jac} is imposed modulo the
quasi-commutativity relation \eqref{qcom}. More precisely, 
following \cite{DSK1} we introduce the subspace
$\mc M_\Delta(R) \subset \Tens(R)[\leq\Delta]$ 
spanned over $\CC$ by all elements of the form
\begin{equation*}
 A \ot \Bigl( \bigl( b\ot c - (-1)^{p(b) p(c)} c\ot b \bigr) \ot D 
 - {:} \Bigl( \int_{-T}^0 \, d\la \, [b_\la c] \Bigr) D {:} \Bigr) ,
\end{equation*}
where
$b,c\in R$, $A,D\in\Tens(R)$ and $\Delta(A \ot b \ot c \ot D) \leq \Delta$.
Then the Jacobi identity \eqref{jac} must hold modulo
$\mc M_\Delta(R)$ for some $\Delta < \Delta(a)+\Delta(b)+\Delta(c)$.

Recall that a vertex algebra $V$ is \emph{strongly generated}
by a subset $\{a^\alpha\} \subset V$ if all monomials 
${:} a^{\alpha_1} \cdots a^{\alpha_s} {:}$ and $\vac$ linearly span $V$.
Here, as usual, a normally ordered product is taken from right to left, and
an empty product is set equal to $\vac$.
A vertex algebra $V$ is \emph{freely generated} 
by an ordered set $\{a^\alpha\} \subset V$ if the monomials
\begin{equation*}
{:} a^{\alpha_1} \cdots a^{\alpha_s} {:} \quad \text{with} \;
\alpha_i \le \alpha_{i+1} \;\text{and}\; \alpha_i < \alpha_{i+1}
\;\text{when}\; p(a^{\alpha_i}) = \bar 1 \quad (1\le i<s) \,,
\end{equation*}
together with $\vac$, form a $\CC$-basis of $V$.

Consider the subspace 
$\mc M(R) = \sum_{\Delta\in \frac12\ZZ_+} \mc M_\Delta(R)$ 
of the tensor algebra $\Tens(R)$. One of the main results of \cite{DSK1} 
is that the $\la$-bracket and Wick product are well defined on the
quotient $V(R) = \Tens(R) / \mc M(R)$ and provide it with the structure of
a vertex algebra. Moreover, $V(R)$ is \emph{freely generated} by 
$R \subset V(R)$, i.e., every ordered $\CC$-basis $\{a^\alpha\}$ of $R$, 
compatible with parity and conformal weight, freely generates $V(R)$.
Conversely, if a vertex algebra $V$ is freely generated by a free 
$\CC[T]$-submodule $R \subset V$
graded by a conformal weight,
then one can endow $R$ with the structure of
a non-linear Lie conformal algebra so that $V \simeq V(R)$
(see \cite{DSK1}).

\subsection{Poisson vertex algebras and non-linear Poisson conformal algebras}

If we remove the integrals (``quantum corrections'') in the axioms
\eqref{qcom}, \eqref{qas} and \eqref{wick} of a vertex algebra,
we arrive at the definition of a \emph{Poisson vertex algebra} 
(cf.\ \cite{FB,DLM2}). 
More precisely, a Poisson vertex algebra is a quintuple
$(\mc V, \vac ,T ,[ \,\cdot\, {}_{\lambda} \,\cdot\, ], \,\cdot\,)$, where
%
$(\mc V,T,[ \,\cdot\, _{\lambda} \,\cdot\, ])$ 
is a Lie conformal superalgebra,
%
$(\mc V,\vac ,T , \,\cdot\,)$ is a unital commutative associative
differential superalgebra,
%
and the operations 
$[ \,\cdot\, {}_{\lambda} \,\cdot\, ]$ 
and $\,\cdot\,$ are
related by the \emph{Leibniz rule} ($=$ commutative Wick formula): 
\begin{equation}\label{xxx1}
[a_{\lambda} (bc) ] = [ a_{\lambda} b ] c
+ (-1)^{p(a)p(b)} b [ a_{\lambda} c ] \,.
\end{equation}
%
%
Here and below, we write the product $a\cdot b$ as simply $ab$.
As for vertex algebras, in a Poisson vertex algebra $\mc V$ 
we can define \emph{$n$-th products} for every $n\in\ZZ$ as follows:
\begin{equation*}
a_{(-n-1)}b = \frac1{n!} (T^n a) b\,, \qquad
a_{(n)}b\ =\ \d_\lambda^n \, [a_\la b]\big|_{\lambda=0}\,,
\qquad n\in\ZZ_+ \,.
\end{equation*}

For example, let 
$\mc V=\CC[u_0,u_1,\dots]$
be the algebra of polynomials in even indeterminates
$u_n$, let $\vac=1$ and $T$ be the derivation of $\mc V$ defined by 
$Tu_n=u_{n+1}$.
Then if we define the $\lambda$-bracket on $\mc V$ by
\begin{equation*}
[P_\la Q] = \sum_{p,q\in\ZZ_+} (-1)^p 
\frac{\partial Q}{\partial u_q} \, (\lambda+T)^{p+q+1}
\frac{\partial P}{\partial u_p} \,,
\end{equation*}
we obtain the so-called 
\emph{Gardner--Faddeev--Zakharov} Poisson vertex algebra.

Given a Lie conformal algebra $R$, one canonically associates to it
the \emph{universal enveloping} Poisson vertex algebra
$\mc V=\Sym(R)$, the symmetric algebra over $R$
as a unital commutative associative differential algebra,
by extending the $\lambda$-bracket of $R$ to $\Sym(R)$
using the Leibniz rule \eqref{xxx1} and skew-symmetry \eqref{ss}.

However, as for vertex algebras, not all Poisson vertex algebras 
$\mc V$ are obtained
as enveloping algebras of Lie conformal algebras and, in general,
the $\lambda$-brackets among generators contain non-linearities.
In order to take into account such non-linearities, 
the notion of a {\it non-linear Poisson conformal algebra} 
is then introduced \cite{DSK2}.
This is a $\CC[T]$-module $R$ together with a $\lambda$-bracket
$R\otimes R\to\CC[\lambda]\otimes \Sym(R)$,
extended to $\Sym(R)$ 
by the Leibniz rule \eqref{xxx1},
which satisfies sesqui-linearity \eqref{ca}, 
skew-symmetry \eqref{ss} and Jacobi identity \eqref{jac}.
It is not hard to check, in a way similar (but much easier) 
to the ``quantum'' case,
that if $R$ is a non-linear Poisson conformal algebra, 
then $\Sym(R)$ has an induced Poisson vertex algebra structure.
Conversely, if $\mc V=\Sym(R)$ 
is a Poisson vertex algebra freely generated by $R$,
then $R$ has a structure of non-linear Poisson conformal algebra.

\subsection{Non-linear deformations of the affine Lie algebra $\what{\sl}_2$}
\label{snldef}

Recall from \cite{K} that the affine Lie algebra $\what{\sl}_2$ 
at level $k\in\CC$ corresponds to the \emph{current Lie conformal algebra} 
$R=\CC[T]\sl_2$, 
in which the $\la$-brackets for the standard generators $\bar h, e, f$
of $\sl_2$ are:
\begin{equation}\label{sl-1}
\begin{split}
[\bar h_\la \bar h] = & 2k\la \,, \qquad
[\bar h_\la e] = 2e \,, \qquad
[\bar h_\la f] = -2f \,, \qquad
\\
& [e_\la f] = \bar h + k\la \,, \qquad
[e_\la e] = [f_\la f] = 0\,.
\end{split}
\end{equation}
We will study ``non-linear'' vertex algebras and Poisson vertex algebras, 
which generalize the above ``linear'' 
one and which include it as a special case.
More precisely, we will keep the above $\la$-brackets the same, 
except that we will allow  $[e_\la f]$ 
to be an arbitrary polynomial in $\la$, $\bar h$ 
and its derivatives. It will be convenient to set $\bar h = 2h$, $k=2\alpha$.
We thus restate the problem as follows.
\begin{problem}\label{p1}
Classify all vertex algebras and Poisson vertex algebras
strongly generated by three elements $h,e,f$ with $\la$-brackets
\begin{equation}\label{e1.1}
\begin{split}
[h_\la h] &= \alpha\la\vac \,, \quad
[h_\la e] = e \,, \quad [h_\la f] = -f \,, \quad
[e_\la e] = [f_\la f] = 0 \,, 
\\ 
[& e_\la f] = {:} P(\la;h,Th,T^2h,\dots) {:} \,,
\end{split}
\end{equation}
where $\alpha\in\CC$ is an unknown constant and 
$P\in\CC[\la,h,Th,T^2h,\dots]$ 
is an unknown polynomial.
We also assume that $h$ is even
while the parities of $e$ and $f$ are not fixed beforehand.
(In the Poisson case, the normally ordered product is replaced by
the commutative associative one.)
\end{problem}

In the vertex algebra case, 
it follows from the commutator formula \eqref{comfla}
that the field $Y(h,z)$ is a \emph{free boson}:
\begin{equation}\label{heis1}
[Y(h,z),Y(h,w)] = \al\,\d_w \delta(z-w) \,.
\end{equation}
Equivalently, the operators $h_{(n)} \in\End V$ satisfy the commutation
relations of the \emph{Heisenberg Lie algebra} with central charge $\al$:
\begin{equation*}
[h_{(m)},h_{(n)}] = \al\, m \delta_{m,-n} \,,
\qquad m,n \in\ZZ \,.
\end{equation*}
The latter also holds in the Poisson case, due to the Jacobi identity
\eqref{jac} and the Leibniz rule \eqref{xxx1}.
Moreover, since $h_{(n)} \vac = 0$ for $n\ge0$, the vacuum vector 
$\vac$ generates a highest weight module $\langle h\rangle$ over
the Heisenberg algebra. For $\al\ne0$, this module is irreducible 
and is known as the \emph{Fock space}; it is unique up to isomorphism
(see e.g.\ \cite[Example 3.5]{K}).

Note that $\langle h\rangle$ can be described as the 
subalgebra of our (Poisson) vertex algebra generated by $h$.
For $\al\ne0$, the linear map
$\CC[h,Th,T^2h,\dots]\to\langle h\rangle$
given by
\begin{equation}\label{inj}
P(h,Th,T^2h,\dots)\,\mapsto\,{:}P(h,Th,T^2h,\dots){:}
\end{equation}
is a vector space isomorphism, which means that $\langle h\rangle$
is freely generated. 
The map \eqref{inj} is well defined, because
by \eqref{qcom2} the elements $T^n h$ 
commute under the normally ordered product.
One should note, however, that by \eqref{qas} the normally ordered product 
in $\langle h\rangle$ is not associative for $\alpha\ne0$; hence, \eqref{inj} 
is not an associative algebra isomorphism unless $\alpha=0$.

If the whole (Poisson) vertex algebra
is freely generated by $h,e,f$,
then it is isomorphic to the universal enveloping (Poisson) vertex algebra
of a non-linear  Lie (respectively, Poisson) 
conformal algebra $R=\CC[T]\langle h,e,f\rangle$ (see \cite{DSK1}).
Hence, in the freely generated case Problem \ref{p1} reduces
to the following easier problem.
\begin{problem}\label{p2}
Classify all 
non-linear Lie conformal algebras and 
non-linear Poisson conformal algebras with three generators $h,e,f$
with $h$ even and $\la$-brackets as in \eqref{e1.1}.
\end{problem}

Let us explain the difference between Problems \ref{p1} and \ref{p2}.
First, if $R$ is a non-linear Lie conformal algebra solving Problem \ref{p2}, 
then its universal enveloping vertex algebra $V(R)$ is freely generated and
solves Problem \ref{p1} with the same $\al$ and $P$. 
The Jacobi identities \eqref{jac} for $V(R)$ can be deduced from
the $\lambda$-brackets \eqref{e1.1} and the other axioms of a vertex algebra.
Moreover, by \cite{DSK1}, any other vertex algebra $V$ 
solving Problem \ref{p1} for the same $\al$ and $P$ is a quotient of $V(R)$.
On the other hand, for certain $\al$ and $P$ 
there exist vertex algebras solving Problem \ref{p1},
which are not quotients of freely generated ones 
(see Section \ref{s-main} below).
These algebras
satisfy the Jacobi identities \eqref{jac} only because of some
additional relations among the generators $h,e,f$ and their derivatives.
In this case, there is no corresponding non-linear Lie conformal algebra 
solving Problem \ref{p2}. Similar remarks apply in the Poisson case.

\subsection{The vertex algebras $V_{\ZZ\sqrt\be}\,$, $V^d_{-1}$ and
$\mc W^{(2)}_n$}
In the present paper we solve Problem \ref{p1} for $\al\ne0$
and Problem \ref{p2} for arbitrary $\al$. Our main classification
results are stated in Section \ref{s-main} below; here we just mention
a few examples.

The first example is provided by the \emph{lattice vertex algebras}
$V_{\ZZ\sqrt\be}\,$, where $\be\in\NN$ (see \cite{B,FLM,K,FB,LL}). 
These algebras solve Problem \ref{p1} with $\al=1/\be$ and 
a homogeneous polynomial $P$ of degree $d=\be-1$ for the grading given by 
$\deg\la=\deg T=\deg h=1$. They are discussed in detail in 
Section \ref{svbe} below.

Another series of vertex algebras solving Problem \ref{p1}
consists of the algebras $V^d_{-1}$ constructed in the present paper.
They correspond to $\alpha=-1$ and an explicit polynomial $P$
of degree $d$ (see Sections \ref{snlie}, \ref{sva1} and \ref{sva2} below).
Since $V^d_{-1}$ is freely generated by $h,e,f$, 
it is the universal enveloping vertex algebra
of a non-linear Lie conformal algebra $R_{-1}^d$
solving Problem \ref{p2}. 

In \cite{FS}, Feigin and Semikhatov introduced a sequence of vertex algebras
$\mc W^{(2)}_n$ ($n\in\NN$). These algebras involve a certain parameter
$k\in\CC$ (similar to the level of the affine algebra) and are only defined 
for $k\ne -n$, because the Virasoro central charge has a pole at $k=-n$
(see \cite[Eq.\ (1.1)]{FS}). For $n=2$, one obtains the 
\emph{affine vertex algebra}
of $\sl_2$ at level $k$, while $\mc W^{(2)}_3$ coincides with the
\emph{Bershadsky--Polyakov $W^{(2)}_3$ algebra}.
The algebras $\mc W^{(2)}_n$ do not solve Problem \ref{p1}, because in them
the $\la$-bracket $[e_\la f]$ is a polynomial not only of $h$ and its
derivatives but also of other elements, which in particular include
the Virasoro one. After comparing our formulas with those in 
\cite[Appendix A]{FS}, one sees that for $n\le 4$
our vertex algebra $V^{n-1}_{-1}$ can be obtained as a certain subquotient
of $\mc W^{(2)}_n$ at the \emph{critical level} $k=-n$.
We believe this is true for all $n$; however, in general
it cannot be done explicitly because the methods of \cite{FS} are 
very different from ours.


\subsection{Plan of the paper}
In Section \ref{s-main}, we state the classification of all 
vertex algebras solving Problem \ref{p1} for $\al\ne0$
and all non-linear Lie conformal algebras solving Problem \ref{p2}
for arbitrary $\al$. The Poisson case is discussed as well.

In Section \ref{s-prel}, we make some preliminary observations
based on grading and change of basis, which
reduce Problems \ref{p1} and \ref{p2} to two equations: 
the Jacobi identity for elements $h,e,f$ 
and the Jacobi identity for $e,e,f$.
The polynomial $P(\la;x_1,x_2,\dots)$ can be assumed homogeneous
of degree $d$, where $\deg\la=1$, $\deg x_k=k$.

We then impose, in Section \ref{s-hef}, 
the Jacobi identity for $h,e,f$.
This determines for every degree $d$ the polynomial 
$P$ explicitly, provided that $\alpha\ne0$.
For $\al=0$, the condition is instead that the polynomial $P$
is independent of $\la$.
The treatment is the same, in that section, both for vertex algebras 
and vertex Poisson algebras.

In Section \ref{s-eef1}, we impose the Jacobi identity for elements $e,e,f$ 
in a vertex algebra solving Problem \ref{p1} with $\alpha\ne0$.
This determines $\alpha$. 
We show that there are two solutions for every degree $d$,
corresponding to $\alpha=-1$ (giving the freely generated $V^d_{-1}$) 
and $\alpha=1/(d+1)$ (giving the lattice vertex algebra $V_{\ZZ\sqrt{d+1}}$).

Next, in Section \ref{s-eef0}, we consider the case 
of a non-linear Lie conformal algebra with $\alpha=0$. We prove that
the only possibility is the current Lie conformal algebra $R=\CC[T]\sl_2$,
thus completing the solution
of Problem \ref{p1} (for $\al\ne0$) and Problem \ref{p2} (for any $\al$)
in the ``quantum'' case.

In Section \ref{s-eefP}, we consider the Poisson, or ``classical,'' case. 
The treatment is
similar to that of the previous two sections, but much simpler.
Besides the current Lie conformal algebra $R=\CC[T]\sl_2$ and
its universal enveloping Poisson vertex algebra, the only solutions
we obtain are with $\alpha=0$ and $P=h^d$.
This provides examples of freely generated Poisson vertex algebras,
which are not ``semiclassical'' limits of any vertex algebra.

In Section \ref{s-wak}, we construct free-field realizations of 
(certain quotients of) the vertex algebras $V^d_{-1}$
solving Problem \ref{p1} with $\alpha=-1$, which generalize the 
Wakimoto realization
of the affine vertex algebra of $\sl_2$ at the critical level $-2$.

In Section \ref{s-zhu}, we determine the Zhu algebra of $V^d_{-1}$, 
which turns out to be one of the associative algebras introduced by 
Smith in \cite{S}. 
The same method is also used to find the Zhu algebra of the
lattice vertex algebra $V_{\ZZ\sqrt{d+1}}\,$, thus reproducing
a result of \cite{DLM1}.

\section{Classification results}\label{s-main}

In this section, we state the classification of all algebras solving 
Problem \ref{p1} for $\al\ne0$ and Problem \ref{p2} for arbitrary $\al$.
First, let us recall a well-known important example, which provided
an early indication that the problems are interesting.

\subsection{Lattice vertex algebras of rank one}\label{svbe}

Fix a positive integer $\be$, and
consider the rank one lattice $\ZZ\sqrt\be\subset\RR$ and the corresponding
\emph{lattice vertex algebra} $V_{\ZZ\sqrt\be}$ 
(see \cite{B,FLM,K,FB,LL}). Let us recall the definition and properties of 
$V_{\ZZ\sqrt\be}\,$, following Sections 5.4 and 5.5 in \cite{K}.
As a vector space, 
\begin{equation*}\label{vbe1}
V_{\ZZ\sqrt\be} = \CC[q,q^{-1};h_{(-1)},h_{(-2)},h_{(-3)},\dots] 
\simeq \langle h\rangle \ot \CC[q,q^{-1}] \,,
\end{equation*}
where $\langle h\rangle=\CC[h_{(-1)},h_{(-2)},\dots]$
is the Fock space for the free boson $Y(h,z)$ 
with $\al=1/\be$ (see \eqref{heis1}).
We let $h_{(n)}$ act trivially on $q^k$ for $n>0$, while $h_{(0)}$
acts on $V_{\ZZ\sqrt\be}$ as $q\d_q$. 
This means that each $q^k$ $(k\in\ZZ)$ 
is a highest weight vector for the Heisenberg algebra;
in particular, the vacuum vector is $\vac=1$.
The parity of $q^k$ is $k\be$ mod $2\ZZ$.

The vertex algebra $V_{\ZZ\sqrt\be}$ is generated by the free boson 
$Y(h,z)$ and by the following quantum fields known as vertex operators
$(k\in\ZZ)$:
\begin{equation}\label{vbe2}
Y(q^k,z) = q^k z^{k\be q\d_q} 
\exp\Bigl( -\sum_{n<0} \frac{z^{-n}}{n} k \be h_{(n)} \Bigr)
\exp\Bigl( -\sum_{n>0} \frac{z^{-n}}{n} k \be h_{(n)} \Bigr) 
\,.
\end{equation}

In fact, $V_{\ZZ\sqrt\be}$ is strongly generated by the elements $h$,
$e=q$, and $f=q^{-1}$. The element $h$ is even, while the parities
of $e$ and $f$ are both equal to $\be$ mod $2\ZZ$.
The $\la$-brackets among the generators $h,e,f$ are given by
\eqref{e1.1}, where $\al=1/\be$ and (cf.\ \cite[Eq.\ (5.5.18)]{K})
\begin{equation}\label{vbe3}
P(\la;h,Th,T^2h,\dots) = \sum_{n=0}^{\be-1} \, 
\frac{\la^n}{n!} \,
S_{\be-1-n} \Bigl(\be\frac{h}{1!} \,,\,
\be\frac{Th}{2!} \,,\,
\be\frac{T^2 h}{3!} \,,\,\dots\Bigr)
\,.
\end{equation}
Here and further,
\begin{equation}\label{vbe4}
S_n(x_1,x_2,x_3,\dots) =
\sum_{\substack{i_1+2i_2+3i_3+\cdots=n \\ i_s \in\ZZ_+}} \,
\frac{x_1^{i_1}}{i_1!} \frac{x_2^{i_2}}{i_2!} \frac{x_3^{i_3}}{i_3!} \cdots
\end{equation}
denotes the \emph{elementary Schur polynomial} of degree $n$
(which is homogeneous when we set $\deg x_k=k$).

Therefore, the vertex algebra $V_{\ZZ\sqrt\be}$ provides a solution
of Problem \ref{p1} with $\al=1/\be$, $\be\in\NN$. However, this
vertex algebra is \emph{not} freely generated by $h,e,f$ 
because of the additional relations
\begin{equation}\label{vbe5}
Te=\be\, {:}he{:} \,, \qquad Tf=-\be\, {:}hf{:} \,,
\end{equation}
which follow from \eqref{vbe2} and the translation covariance axiom.
Finally, note that $V_{\ZZ\sqrt\be}$ is a \emph{simple} vertex algebra,
i.e., it does not have nontrivial proper ideals.

\subsection{Non-linear Lie conformal algebras $R^d_{-1}$}\label{snlie}

Consider the ring $\CC[\la,h,Th,$ $T^2h,$ $\dots]$,
equipped with the derivation $T$ such that $T\la=0$.
For an arbitrary polynomial 
$p(\la)\in\CC[\la]$, we let
\begin{equation}\label{nlie1}
P(\la;h,Th,\dots) = p(\lambda+T-h) \, 1
\in \CC[\la,h,Th,\dots] \,,
\end{equation}
where $p(\lambda+T-h)$ is expanded by letting $T$ act on its right and by letting $T1=0$.
So for example,
\begin{equation*}
(\la+T-h)^2 \, 1 = (\la+T-h)(\la-h) = \la^2-2\la h-Th+h^2\,.
\end{equation*}
When we consider $P$ under the normally ordered product, 
it becomes a polynomial in $\la$ with coefficients belonging to 
a non-linear Lie conformal algebra or a vertex algebra. 
In particular, we let ${:}1{:}=\vac$ be the vacuum vector.

\begin{proposition}\label{pnlie}
For every polynomial $p\in\CC[\la]$, 
there exists a non-linear Lie conformal algebra $R=\CC[T]\langle h,e,f\rangle$
with the $\lambda$-brackets \eqref{e1.1}, 
where $h$, $e$ and $f$ are even, $\al=-1$ and 
$P(\lambda;h,Th,\dots)=p(\lambda+T-h)1$.
\end{proposition}

The proof will be given in Section \ref{spfal-1} below.

We will show in Section \ref{sgrsym} that, without loss of generality, 
one can assume that the polynomial $P$ in Problem \ref{p2} is homogeneous 
for the grading given by $\deg\la=\deg T=\deg h=1$.
If $P$ is homogeneous of degree $d$ and is given by \eqref{nlie1}, then
up to rescaling $p(\la)=\la^d$. In this case, we will denote the 
non-linear Lie conformal algebra $R$ from the proposition as $R_{-1}^d$.
Note that $d=1$ corresponds to $[e_\la f]=\la-h$, which after rescaling
gives the current Lie conformal algebra $\CC[T]\sl_2$ 
at the critical level $k=-2$ (see \eqref{sl-1}).

\subsection{Solution of Problem \ref{p2} in the ``quantum" case.}\label{sva1}

The following theorem, which will be proved in 
Sections \ref{spfalnz} and \ref{spfalz}
below, is one of the main results of the paper.
\begin{theorem}\label{th-june12}
A complete classification of non-linear Lie conformal algebras 
$R=\CC[T]\langle h,e,f\rangle$ 
with $h$ even and with $\lambda$-bracket as in \eqref{e1.1} is the following.
Assume, without loss of generality, that the polynomial $P$ in \eqref{e1.1} 
is homogeneous of degree $d\geq1$ {\rm(}with respect to the grading given by 
$\deg\la=\deg T=\deg h=1)$.
Then{\rm:}
\begin{enumerate}
\item
When $d=1$, $\alpha$ is arbitrary, and $R$ 
is isomorphic to the current Lie conformal
algebra $\CC[T]\sl_2$ at level $k=2\alpha$, 
i.e., after rescaling
\begin{equation*}
[e_\la f]=h+\alpha\lambda\,.
\end{equation*}

\item 
When $d\geq2$, $\alpha=-1$, and $R$ is isomorphic to the
non-linear Lie conformal algebra $R_{-1}^d$, i.e., $e,f$ are even and
\begin{equation}\label{nlie5}
[e_\la f]={:}(\lambda+T-h)^d \, 1{:}\,.
\end{equation}
\end{enumerate}
\end{theorem}

In Section \ref{s-hef}, we will provide another
formula for the $\lambda$-bracket in $R_{-1}^d$, which involves the 
elementary Schur polynomials \eqref{vbe4} and is similar to
formula \eqref{vbe3} above.

\subsection{Solution of Problem \ref{p1} in the ``quantum" case for 
$\alpha\neq0$.}\label{sva2}

The following result describes all vertex algebras solving Problem \ref{p1}
with $\al\ne0$ and a homogeneous polynomial $P$.
\begin{theorem}\label{th-june12b}
Consider vertex algebras strongly generated by elements 
$h,e,f$ with $\lambda$-bracket relations as in \eqref{e1.1},
where $\al\ne0$ and the polynomial $P$ is homogeneous of degree $d\geq1$
{\rm(}when $\deg\la=\deg T=\deg h=1)$. 
Then, up to isomorphism,
a complete list of such vertex algebras $V$ is{\rm:}

\begin{enumerate}
\item
When $d=1$, $\alpha$ is arbitrary, and $V$ is a quotient 
of the universal enveloping  vertex algebra of the current Lie 
conformal algebra $\CC[T]\sl_2$ at level $k=2\alpha$.

\item
When $d\geq2$, $\alpha=-1$, and $V$ is a quotient 
of the universal enveloping vertex algebra $V^d_{-1}:=V(R_{-1}^d)$
of the non-linear Lie conformal algebra $R_{-1}^d$.

\item 
When $d\geq2$, $\alpha=1/(d+1)$, and $V$ is isomorphic to
the lattice vertex algebra $V_{\ZZ\sqrt{d+1}}\,$.
In this case, $e$ and $f$ must have the same parity as $d+1$.
\end{enumerate}
\end{theorem}

We will prove this theorem in Section \ref{spfalnz} below.
The arguments in the proof do not work for $\alpha=0$, and in fact
they may be used to construct counterexamples.

Notice that, by the Frenkel--Kac construction \cite{FrK,K}, 
the lattice vertex algebra $V_{\ZZ\sqrt2}$ is a quotient of 
the universal enveloping  vertex algebra of the current Lie 
conformal algebra $\CC[T]\sl_2$ at level $k=1$, which corresponds
to $\al=1/2$.
Thus, for $d=1$, part (c) of Theorem \ref{th-june12b}
is included in part (a).

\subsection{Solution of Problems \ref{p1} and \ref{p2} in the Poisson case}
\label{spc}

It is easy to check that, for any polynomial $p(\la)\in\CC[\la]$,
the $\lambda$-bracket 
\begin{equation*}
[e_\la f]=p(h) \,,
\end{equation*}
together with the other formulas in \eqref{e1.1} with $\al=0$, 
provides an example of a non-linear Poisson conformal algebra 
solving Problem \ref{p2}. The following two theorems will be proved in 
Section \ref{s-eefP}.
\begin{theorem}\label{th-pc1}
A complete classification of non-linear Poisson conformal algebras
$R=\CC[T]\langle e,f,h\rangle$ with $h$ even and $\lambda$-bracket 
as in \eqref{e1.1} is the following{\rm:}

\begin{enumerate}
\item
The current Lie conformal algebra $R=\CC[T]\sl_2$ at level $k=2\al$.
In this case, $e$ and $f$ are even and the polynomial $P$ is homogeneous
of degree $1$.

\item
For $\al=0$, both $e$ and $f$ are even, and $P=p(h)$ 
is an arbitrary polynomial of $h$.
\end{enumerate}
\end{theorem}

\begin{theorem}\label{t4.1}
Let $\mc V$ be a Poisson vertex algebra
strongly generated by elements $h,e,f\in\mc V$ with $\lambda$-bracket relations
as in \eqref{e1.1} with $\alpha\neq0$.
Then $\mc V$ is a quotient of the universal enveloping 
Poisson vertex algebra $\Sym(R)$
of the current Lie conformal algebra $R=\CC[T]\sl_2$ at level $k=2\alpha$.
\end{theorem}

\section{Preliminary observations}\label{s-prel}

First of all, we note that 
\eqref{e1.1} does not provide explicitly all possible $\la$-brackets
among the generators $h,e,f$, but the remaining ones are determined 
by the skew-symmetry axiom \eqref{ss}:
\begin{equation*}
[e_\la h] = -e \,, \quad [f_\la h] = f \,, \quad 
[f_\la e] = -(-1)^{p(e)p(f)} \, {:} P(-\la-T;h,Th,T^2h,\dots) {:} \,.
\end{equation*}
Then the sesqui-linearity \eqref{ca} is used to extend the $\la$-bracket
to $\CC[T]\langle h,e,f\rangle$.
Hence, the sesqui-linearity and skew-symmetry are automatically satisfied, 
and we only need to impose the Jacobi identities \eqref{jac} 
involving the generators $h,e$ and $f$.
Furthermore, the only Jacobi identities that are not  
trivially satisfied are the ones for the triples 
$(h,e,f)$, $(e,e,f)$, and $(f,f,e)$.

\subsection{Grading and symmetry conditions}\label{sgrsym}
In order to solve Problem \ref{p2}, we need to find all $\al$ and $P$
such that the $\la$-bracket given by \eqref{e1.1}
satisfy the Jacobi identities for the triples 
$(h,e,f)$, $(e,e,f)$, and $(f,f,e)$.
These identities lead to certain equations, which are linear in $P$.
Therefore, for fixed $\al\in\CC$, the set of all polynomials $P$
solving Problem \ref{p2} forms a complex vector space.
(This conclusion does \emph{not} hold for the set of polynomials $P$
solving Problem \ref{p1}, because each particular solution may involve
its own set of additional relations among the generators; see the discussion
at the end of Section \ref{snldef}.)

Hence, for Problem \ref{p2},
we can assume that $P$ is a \emph{homogeneous} polynomial
of degree $d$ for the grading given by $\deg\la=\deg T=\deg h=1$.
Then the $\lambda$-bracket relations \eqref{e1.1} are homogeneous as well.
This grading is compatible with a grading by \emph{conformal weight}
in such a way that (cf.\ \eqref{wtla}):
\begin{equation}\label{e1.2}
\Delta(h)=1\,, \qquad \Delta(e)+\Delta(f)=d+1 \,.
\end{equation}
We are going to make this assumption for Problem \ref{p1} as well.
Notice also that, by replacing the generator $e$ by $\gamma e$
($\gamma\in\CC$), one obtains another solution with the polynomial
$P$ replaced by $\gamma P$.

Next, it follows from the skew-symmetry \eqref{ss} that relations
\eqref{e1.1} are invariant under the change of generators
\begin{equation}\label{symef}
\tilde h= -h \,, \qquad \tilde e= f \,, \qquad \tilde f= e \,,
\end{equation}
if we replace the polynomial $P$ by
\begin{equation}\label{ptilde}
\tilde P(\lambda;h,Th,T^2h,\dots) 
= -(-1)^{p(e)p(f)} \, P(-\lambda-T;-h,-Th,\dots)\,.
\end{equation}
Hence, if $R=\CC[T]\langle h,e,f\rangle$ is a non-linear Lie conformal algebra
satisfying the assumptions of Problem \ref{p2} for a given choice of $\alpha$ and $P$,
then $\tilde R=\CC[T]\langle \tilde f,\tilde h,\tilde e\rangle$ is also a non-linear
Lie conformal algebra of the same type, with $\tilde \alpha=\alpha$
and $\tilde P(\lambda;h,Th,\dots)$ as in \eqref{ptilde}.
Conversely, if the Jacobi identities for the triples $(h,e,f)$ and $(e,e,f)$
hold with both polynomials $P$ and $\tilde P$,
then the Jacobi identity for the triple $(f,f,e)$ follows automatically.
In particular,
if $P$ satisfies the \emph{symmetry condition} 
$\tilde P=\pm P$, then the Jacobi identity for the triple
$(f,f,e)$ follows from the one for $(e,e,f)$.
In fact,
we will show in Section \ref{s-hef} below that for $\al\ne0$
this symmetry condition follows from the Jacobi identity 
for $(h,e,f)$.


In conclusion, for both Problem \ref{p1} and \ref{p2},
we will assume that $P$ is homogeneous of degree $d$, and if
$\tilde P=\pm P$, we only need to impose the Jacobi identities for the 
triples $(h,e,f)$ and $(e,e,f)$. This will be done separately 
in the following sections. 

\subsection{Technical results}\label{stech}

Here, we collect several computational results
which will be useful in the sequel. 
Throughout this subsection $\be$ will be a fixed complex number.
For a formal series $\ph(x)\in\CC[[x,x^{-1}]]$, we will denote by 
$\Reg_x \ph(x)$ its \emph{regular part}, i.e., 
\begin{equation*}
\Reg_x x^n\,=\,
\left\{\begin{array}{rl}
\!\! x^n, & \text{ if }\, n\geq0\,,\\
\!\! 0\;, & \text{ if }\, n<0\,.
\end{array}\right.
\end{equation*}
Let $\Delta_x$ be the \emph{difference operator} acting on
regular series by
\begin{equation*}
(\Delta_x \ph)(x) = \frac{\ph(x)-\ph(0)}x \;, \qquad \ph(x)\in\CC[[x]] \,.
\end{equation*}

\begin{lemma}\label{ldelreg}
We have{\rm:}
\begin{equation*}
\Delta_x \Reg_x \ph(x) = \Reg_x x^{-1} \ph(x) 
\,, \qquad \ph(x)\in\CC[[x,x^{-1}]] \,.
\end{equation*}
\end{lemma}
\begin{proof}
It suffices to check this equation for $\ph(x)=x^n$ $(n\in\ZZ)$,
in which case it is straightforward.
\end{proof}

We introduce the hypergeometric function
\begin{equation*}
\Phi_\be(x) = \sum_{n=0}^\infty \binom{\be}n \frac{x^n}{n!} \;,
\end{equation*}
where, as usual, the binomial coefficient is given by
$\binom{\be}n = \frac{\be(\be-1)\cdots(\be-n+1)}{n!}$.
Using that $(n+1)\binom{\be}{n+1} = (\be-n)\binom{\be}n$,
it is easy to check that $\Phi_\be(x)$ satisfies the
differential equation
\begin{equation}\label{fbe2}
\bigl( x\d_x^2 + (x+1)\d_x -\be \bigr) \Phi_\be(x) = 0 \,.
\end{equation}
Let us also introduce the formal series
\begin{equation}\label{fbe3}
\Psi_\be(x,y) 
= \d_y \Phi_\be(-xy)
= \sum_{n=1}^\infty \binom{\be}n \frac{(-x)^n y^{n-1}}{(n-1)!} \;.
\end{equation}

\begin{proposition}\label{pro-june12}
Let\/ $a$ and\/ $b$ be two elements in a vertex algebra satisfying\/ 
$[b_\la b] \in\CC[\la]\vac$ and\/ $[b_\la a]=\be a$, where $b$ is even.
Then the following identities hold{\rm:}
\begin{align}
\label{june12-c}
{:} (e^{yT} a) (e^{x(T+b)} 1) {:}
&= \Reg_y \, (1-xy^{-1})^\be \,
{:} (e^{x(T+b)} 1) \, (e^{yT} a) {:} \,,
\\ \label{june12-e}
{:} a \, \big(e^{x(T+b)} 1\big) {:}
&= {:} (e^{x(T+b)} 1) \, \big(\Phi_\be(-xT) a\big) {:} \,,
\\ \label{june12-d}
\bigl[ a_\la {:} \big(e^{x(T+b)} 1\big) {:} \bigr] 
&= {:} (e^{x(T+b)} 1) \, \big(\Psi_\be(x,\lambda+T) a\big) {:} \,,
\end{align}
where $x$ and $y$ are formal variables and\/ $\be\in\CC$.
In the right-hand side of \eqref{june12-c},
the function $(1-xy^{-1})^\be$ is expanded as a power series of\/ $x,y$ 
in the domain $|x|<|y|$.
\end{proposition}

Recall that a normally ordered product of more than two factors 
is taken from right to left (cf.\ \eqref{mono2}),
and that $T1=0$. In the right-hand sides of the above equations,
$e^{x(T+b)} 1$ is considered as a formal power series in $x$ whose
coefficients are polynomials in $b,Tb,T^2 b$, etc. The latter are
multiplied by elements of the form $T^k a$, and give elements 
of the vertex algebra only after taking the normally ordered product.
This is well defined because, by \eqref{qcom2}, the condition 
$[b_\la b] \in\CC[\la]\vac$
implies that all elements $T^k b$ $(k\in\ZZ_+)$ 
commute under the normally ordered product.

So, for example, we have
\begin{equation*}
{:}\bigl((T+b)^21\bigr)a{:} = {:}(Tb)a{:}+{:}b({:}ba{:}){:} \,,
\end{equation*}
which differs from 
\begin{equation*}
{:}\bigl({:}(T+b)^21{:}\bigr)a{:}={:}(Tb)a{:}+{:}({:}bb{:})a{:} \,,
\end{equation*}
or from 
\begin{equation*}
{:}(T+b)^2a{:}={:}(Tb)a{:}+{:}b({:}ba{:}){:}+2{:}b(Ta){:}+T^2a \,.
\end{equation*}

\begin{proof}[Proof of Proposition \ref{pro-june12}]
To prove \eqref{june12-c}, we notice that both sides are equal to
$e^{yT} a$ for $x=0$, and we are going to show they both satisfy
the same first-order differential equation in $x$. 
Denote the left-hand side by $A(x,y)$.
Using \eqref{qcom2} and the fact that $T$ 
is a derivation of the normally ordered product, we find:
\begin{align*}
(\d_x+\d_y-T) & A(x,y) = {:} (e^{yT} a) \, b \, \big(e^{x(T+b)} 1\big) {:}
\\
&= {:} b \, (e^{yT} a) \, \big(e^{x(T+b)} 1\big) {:}
+ {:} \Bigl( \int_{-T}^0 \, d\la \, [(e^{yT} a)_\la b] \Bigr) 
\big(e^{x(T+b)} 1\big) {:} \,.
\end{align*}
Then, by the sesqui-linearity \eqref{ca} and skew-symmetry \eqref{ss},
\begin{equation*}
\int_{-T}^0 \, d\la \, [(e^{yT} a)_\la b] 
= \int_{-T}^0 \, d\la \, e^{-y\la} [a_\la b] 
= -\int_{-T}^0 \, d\la \, e^{-y\la} \be a 
= - \be \, \frac{e^{yT}-1}y \, a \,.
\end{equation*}
Putting these together, we obtain:
\begin{equation*}
(\d_x+\d_y-T) A(x,y) = {:} b A(x,y) {:} -\be\Delta_y A(x,y) \,.
\end{equation*}
The right-hand side of \eqref{june12-c} 
satisfies the same differential equation, due to Lemma \ref{ldelreg}
and the fact that
\begin{equation*}
(\d_x+\d_y) (1-xy^{-1})^\be = -\be y^{-1} (1-xy^{-1})^\be \,.
\end{equation*}
This proves \eqref{june12-c}.

To derive \eqref{june12-e}, we set $y=0$ in both sides of 
formula \eqref{june12-c}. Using the binomial expansion in the domain $|x|<|y|$,
it is straightforward to compute
\begin{equation*}
\Bigl( \Reg_y \, (1-xy^{-1})^\be \, e^{yT} \Bigr)\Big|_{y=0}
= \sum_{n=0}^\infty \binom{\be}n (-x)^n \frac{T^n}{n!}
= \Phi_\be(-xT) \,.
\end{equation*}

To prove \eqref{june12-d}, we are going to follow the same strategy
as above. We first note that both sides vanish for $x=0$.
Denoting the left-hand side by $B(x,\la)$, we compute
\begin{equation*}
\d_x B(x,\la) = \bigl[ a_\la {:} (T+b) \big(e^{x(T+b)} 1\big) {:} \bigr] 
= (\la+T) B(x,\la) + \bigl[ a_\la {:} b \, \big(e^{x(T+b)} 1\big) {:} \bigr] \,,
\end{equation*}
using the sesqui-linearity \eqref{ca}. By the noncommutative Wick formula
\eqref{wick} and $[a_\la b]=-\be a$, we have:
\begin{equation*}
\bigl[ a_\la {:} b \, \big(e^{x(T+b)} 1\big) {:} \bigr]
= - \be {:} a \, \big(e^{x(T+b)} 1\big) {:} + {:} b \, B(x,\la) {:}
- \be \int_0^{\la} \, d\mu \, B(x,\mu) \,.
\end{equation*}
Therefore, $B(x,\la)$ satisfies the following differential 
equation in $x$:
\begin{equation*}
\d_x B(x,\la) = {:} (\la+T+b) \, B(x,\la) {:}
- \be \int_0^{\la} \, d\mu \, B(x,\mu)
 - \be \, {:} (e^{x(T+b)} 1) \, \big(\Phi_\be(-xT) a\big) {:} \,.
\end{equation*}
For the last term we used \eqref{june12-e}.
To show that the right-hand side of \eqref{june12-d} 
satisfies the same differential equation, it suffices to check that
\begin{equation*}
\d_x \Psi_\be(x,\lambda+T) = (\la+T) \Psi_\be(x,\lambda+T)
- \be \int_0^{\la} \, d\mu \, \Psi_\be(x,\mu+T)
- \be \Phi_\be(-xT) \,.
\end{equation*}
Using that $\Psi_\be(x,\lambda+T) = \d_\lambda \Phi_\be(-x(\lambda+T))$,
it is easy to reduce this identity to
$$
\partial_x\partial_\mu \Phi_\beta(-x\mu) = \mu\partial_\mu \Phi_\beta(-x\mu)
-\beta\Phi_\beta(-x\mu)\,,
$$
which holds thanks to equation \eqref{fbe2}.
This completes the proof.
\end{proof}

Proposition \ref{pro-june12} takes a particularly nice form when $\be=\pm1$, 
since 
\begin{equation}\label{fbe4}
\Phi_1(x) = 1+x \,, \; \Psi_1(x,y) = -x \,, \qquad 
\Phi_{-1}(x) = e^{-x} \,, \; \Psi_{-1}(x,y) = x e^{xy} \,. 
\end{equation}

\begin{corollary}\label{clabe}
Under the assumptions of Proposition \ref{pro-june12}, 
for any polynomial $p(\lambda)\in\CC[\la]$, we have{\rm:}
\begin{enumerate}
\item[(a)] For $\beta=1$,
\begin{align*}
{:}a\big(p(T+b)1\big){:} &= {:}\big(p(T+b)1\big)a{:}
-{:}\big(p^\prime(T+b)1\big)(Ta){:}\,, \\
\big[a _\lambda {:}\big(p(T+b)1\big){:}\big] &= 
-{:}\big(p^\prime(T+b)1\big)a{:}\,,
\end{align*}
\item[(b)] For $\beta=-1$,
\begin{align*}
{:}a\big(p(T+b)1\big){:} &= {:}p(T+b)a{:}\,, \\
\big[a _\lambda {:}\big(p(T+b)1\big){:}\big] &= {:}\big(p^\prime(\lambda+T+b)1\big)a{:}\,,
\end{align*}
\end{enumerate}
where $p'$ denotes the derivative of\/ $p$.
\end{corollary}

One important difference between the above formulas is that in the case
$\be=1$ the translation operator $T$ gives $0$ when applied to $1$
and it does not carry to the element $a$,
while in the case $\be=-1$ the operator $T$ is applied to $a$
(recall the observations after Proposition \ref{pro-june12}).

\begin{proof}[Proof of Corollary \ref{clabe}]
By \eqref{june12-d} and \eqref{fbe4}, we have in the case $\be=1$:
\begin{align*}
{:}a\big(e^{x(T+b)}1\big){:} &= {:}\big(e^{x(T+b)}1\big)(a-xTa){:}\,, \\
\bigl[ a_\la {:} \big(e^{x(T+b)} 1\big) {:} \bigr] 
&= - x \, {:} (e^{x(T+b)} 1) \, a {:} \,.
\end{align*}
Similarly, in the case $\be=-1$:
\begin{align*}
{:}a\big(e^{x(T+b)}1\big){:} &= {:}e^{x(T+b)}a{:}\,, \\
\bigl[ a_\la {:} \big(e^{x(T+b)} 1\big) {:} \bigr] 
&= x \, {:} e^{x(\la+T+b)} a {:}\,,
\end{align*}
using that $T$ is a derivation of the normally ordered product.
The corollary now follows by comparing the coefficients of the various
powers of $x$ and by linearity.
\end{proof}
%

\section{Jacobi identity for $h,e,f$}\label{s-hef}

Throughout this section, we will work in a vertex algebra or
a Poisson vertex algebra solving Problem \ref{p1}.
We will consider the Jacobi identity for the triple $(h,e,f)$:
\begin{equation}\label{e3.1}
[h_\la [e_\mu f]] = [[h_\la e]_{\la+\mu} f] + [e_\mu [h_\la f]]
\end{equation}
as an equation for the polynomial $P$. We will show that
for $\al\ne0$ this equation determines $P$ uniquely up to a 
multiplicative constant for each degree $d$.

\subsection{Equations for the polynomial $P$}\label{seqp}

By \eqref{e1.1}, the right-hand side of \eqref{e3.1} is simply
\begin{equation*}
[e_{\la+\mu} f] - [e_\mu f] = 
{:}P(\la+\mu;h,Th,T^2h,\dots){:}
-{:}P(\mu;h,Th,T^2 h,\dots){:} \,.
\end{equation*}
To compute the left-hand side of \eqref{e3.1}, we observe that
the Wick formula \eqref{wick} when applied to polynomials of $T^k h$
reduces to its \emph{commutative} version \eqref{xxx1},
because $h$ is a free field, i.e., $[h_\lambda h]$ is annihilated by $T$. 
We obtain that
\begin{equation*}
\begin{split}
[h_\la {:}P(\mu;h,Th,T^2h,\dots){:}] 
&= \sum_{k=1}^\infty \, [h_\la (T^{k-1} h)] \;
{:} \frac{\d P}{\d x_k}(\mu;h,Th,T^2h,\dots){:}
\\
&= \sum_{k=1}^\infty \al\la^k \, 
{:} \frac{\d P}{\d x_k}(\mu;h,Th,T^2h,\dots){:} \,,
\end{split}
\end{equation*}
where $x_k = T^{k-1} h$ $(k\in\NN)$.
The above formulas remain the same in the Poisson case
if we replace the normally ordered product by the commutative associative one.

Since the map \eqref{inj} is injective,
we can ignore the normal orderings in the above two equations; then
\eqref{e3.1} becomes equivalent to the following identity in the polynomial 
ring $\CC[\la,\mu,x_1,x_2,\dots]$:
\begin{equation*}
P(\la+\mu;x_1,x_2,\dots) 
- P(\mu;x_1,x_2,\dots)
= \sum_{k=1}^\infty \, \alpha \la^k \, 
\frac{\d P}{\d x_k}(\mu;x_1,x_2,\dots) \,.
\end{equation*}
By Taylor's formula, this identity is equivalent to the system of equations
\begin{equation}\label{e3.5}
\frac{1}{k!} \, \frac{\d^k P}{\d\la^k} (\la;x_1,x_2,\dots) 
= \alpha \, \frac{\d P}{\d x_k} (\la;x_1,x_2,\dots) \,,
\qquad k=1,2,\dots \;.
\end{equation}
{}

\subsection{Solving the equations}\label{ssoleq}

For $\alpha=0$, equations \eqref{e3.5} hold if and only if
$P(\la;x_1,x_2,\dots)$ 
is independent of $\la$.
Until the end of this section, we will assume that $\alpha\ne0$ 
and we set $\beta=1/\alpha$.

Equations \eqref{e3.5} are homogeneous if we define a
grading by $\deg\la=1$, $\deg x_k = k$. 
After the substitution $x_k = T^{k-1} h$, this grading agrees with
the one defined earlier (in Section \ref{sgrsym}) by
$\deg\la=\deg T=\deg h=1$. Henceforth, we will assume that
the polynomial $P$ is homogeneous of degree $d\ge1$.

\begin{lemma}\label{l-july16}
For $\al\ne0$, every solution $P$ of equations \eqref{e3.5}, 
homogeneous of degree $d$ with respect to the grading\/ 
$\deg\la=1$, $\deg x_k=k$, is of the form
\begin{equation}\label{e3.9}
P(\la;x_1,x_2,x_3,\dots) 
= \gamma\, S_d\Bigl(\la+\be\frac{x_1}{1!} \,,\,
\be\frac{x_2}{2!} \,,\, \be\frac{x_3}{3!} \,,\, \dots\Bigr) \,,
\end{equation}
where $\be=1/\al$, $\gamma\in\CC$, and
$S_d$ is the elementary Schur polynomial defined by \eqref{vbe4}.
\end{lemma}
\begin{proof}
After making the change of variables
\begin{equation*}
y_k = \be \frac{x_k}{k!} \,, \qquad 
x_k = \al \, k! \, y_k \,, \qquad k=1,2,\dots \,,
\end{equation*}
equations \eqref{e3.5} can be rewritten as follows:
\begin{equation*}
\frac{\d^k P}{\d\la^k} = \frac{\d P}{\d y_k}
\,, \qquad k=1,2,\dots \;.
\end{equation*}
For $k=1$, this implies that
\begin{equation*}
P(\la;x_1,x_2,\dots) = Q(\la+y_1,y_2,y_3,\dots)
\end{equation*}
for some polynomial $Q$.
The rest of the equations are then equivalent to:
\begin{equation}\label{e3.8}
\frac{\d Q}{\d y_k} (y_1,y_2,\dots) =
\frac{\d^k Q}{\d y_1^k} (y_1,y_2,\dots)
\,, \qquad k=1,2,\dots \;.
\end{equation}
If the coefficient of $y_d$ in $Q(y_1,y_2,\dots)$ 
is equal to $\gamma$, then
\begin{equation*}
\frac{\d^{i_1}}{\d y_1^{i_1}}
\frac{\d^{i_2}}{\d y_2^{i_2}} \cdots
\frac{\d^{i_d}}{\d y_d^{i_d}} Q 
= \frac{\d Q}{\d y_d} 
= \gamma \,,
\end{equation*}
for any choice of $i_s\in\ZZ_+$ such that
$1 i_1 + 2 i_2 + \dots + d i_d = d$.
Therefore, the coefficient of $y_1^{i_1} \cdots y_d^{i_d}$ in 
$Q(y_1,y_2,\dots)$ is equal to $\gamma / i_1! \cdots i_d! \,$,
and hence $Q=\gamma S_d$.
\end{proof}

Recall that the generating function of the elementary Schur polynomials is
\begin{equation}\label{e3.10}
\sum_{n=0}^\infty z^n S_n (y_1,y_2,\dots)
= \exp \Bigl( \sum_{k=1}^\infty z^k y_k \Bigr) \,,
\end{equation}
where $S_0 \equiv 1$.
Indeed, it is obvious that the function
$\exp ( \sum z^k y_k )$ satisfies equations~\eqref{e3.8}.
{}From \eqref{e3.9} and \eqref{e3.10}, we obtain
\begin{equation}\label{e3.13}
\begin{split}
P(\la;h,& Th,\dots)
= \gamma \Res_z z^{-d-1} \exp \Bigl(z\la+ \sum_{k=1}^\infty 
z^k \beta \frac{T^{k-1}h}{k!} \Bigr)
\\
&= \gamma \Res_z z^{-d-1} e^{z\la}
\exp \Bigl(\frac{e^{zT}-1}{T} \be h \Bigr) \,.
\end{split}
\end{equation}

\begin{remark}\label{r-july17}
Expanding the exponential $e^{z\la}$ in \eqref{e3.13} and taking the
residue, we get
\begin{equation*}
P(\la;h,Th,T^2h,\dots) = \gamma\, \sum_{n=0}^d \, 
\frac{\la^n}{n!} \,
S_{d-n} \Bigl(\be\frac{h}{1!} \,,\,
\be\frac{Th}{2!} \,,\,
\be\frac{T^2 h}{3!} \,,\,\dots\Bigr)
\,.
\end{equation*}
This is exactly formula \eqref{vbe3} for $\be=d+1$ and $\gamma=1$.
Conversely, \eqref{vbe3} can be rewritten as \eqref{e3.9} with $\be=d+1$.
\end{remark}

\subsection{Symmetry condition}

Now we will show that the polynomial $P$ in Lemma \ref{l-july16} 
satisfies the symmetry condition of Section \ref{sgrsym}.

\begin{lemma}\label{l-july17-1}
The polynomial $P$, given by \eqref{e3.9}, satisfies 
$$
P(-\lambda-T;h,Th,\dots) = (-1)^d \, P(\lambda;h,Th,\dots)\,.
$$
Equivalently, we have $\tilde P=(-1)^{d+1+p(e)p(f)}P$, where $\tilde P$ is defined by \eqref{ptilde}.
\end{lemma}
\begin{proof}
Using \eqref{e3.13} and the fact that $T$ is a derivation, we compute:
\begin{align*}
P(-\lambda-T &; -h,-Th,\dots)
= \gamma \Res_z z^{-d-1} e^{z(-\la-T)}
\exp \Bigl(\frac{e^{zT}-1}{T} \be (-h) \Bigr)
\\
&= \gamma \Res_z z^{-d-1} e^{-z\la}
\exp \Bigl(\frac{e^{-zT}-1}{T} \be h \Bigr)
\\
&= (-1)^d \, P(\la;h,Th,\dots) \,,
\end{align*}
as claimed.
\end{proof}

\subsection{Another formula for the polynomial $P$}\label{sanfor}

Equation \eqref{e3.13} can also be written as
\begin{equation*}
P(\la;h,Th,\dots) = \gamma \Res_z z^{-d-1} e^{z\la}
\exp \Bigl(\be\int_0^z d x\ e^{xT}h \Bigr) \,.
\end{equation*}
Now we will give a formula equivalent to it.

\begin{lemma}\label{l-july17-2}
The following identity holds in the ring\/
$\CC[h,Th,T^2h,\dots][[z]]${\rm:}
\begin{equation*}
\exp\Bigl(\beta \int_0^z d x \, e^{xT} h\Bigr) 
= e^{z(T+\beta h)} \, 1 \,,
\end{equation*}
where $\beta\in\CC$ and in the right-hand side $T1=0$.
\end{lemma}
\begin{proof}
Denote the left-hand side of this equation by $A(z)$.
Since $A(0)=1$, it is enough to check that $A(z)$ satisfies the
differential equation $dA/dz = (T+\beta h) A(z)$. Using that $T$ is a 
derivation, we find
\begin{equation*}
T A(z)
= \beta \Bigl(\int_0^z d x \, e^{xT} Th\Bigr) A(z)
= \beta \bigl( (e^{zT}-1) h \bigr) A(z) \,.
\end{equation*}
On the other hand, $dA/dz = \beta(e^{zT} h) A(z)$,
which completes the proof.
\end{proof}

Combining Lemmas \ref{l-july16}, \ref{l-july17-1} and \ref{l-july17-2}, we get the main
result of this section.

\begin{proposition}\label{p-apr12-1}
In any vertex algebra or Poisson vertex algebra solving Problem \ref{p1},
with a homogeneous polynomial $P$ of degree $d\ge1$,
one of the following two possibilities holds{\rm:}
\begin{enumerate}
\item
$\al=0$ and the polynomial $P(\la;h,Th,\dots)$ is independent of\/ $\la$.

\item
$\al\ne0$, $\be=1/\al$, and for some $\delta\in\CC${\rm:}
\begin{equation}\label{e-apr12-3}
P(\la;h,Th,\dots) = \delta\, (\la+T+\beta h)^d \, 1 \,.
\end{equation}
\end{enumerate}
Conversely, given \eqref{e1.1}, if either {\rm(a)} or {\rm(b)} holds,
then the Jacobi identity for the triple $(h,e,f)$ is satisfied.
Moreover, if {\rm(b)} holds, then the Jacobi identity for the triple $(f,f,e)$
follows from the Jacobi identity for the triple $(e,e,f)$.
\end{proposition}

\begin{remark}\label{r-apr12-2}
The above results can be reformulated in terms of 
the elementary Schur polynomials as follows. 
Let $T$ be the derivation of the polynomial ring 
$\CC[y_1,y_2,\dots]$ defined by $Ty_k = (k+1) y_{k+1}$. Then one has
\begin{equation*}
S_n(y_1,y_2,\dots) = \frac1{n!} (y_1+T)^n \, 1 \,,
\end{equation*}
which can be derived from the recursive relation
\begin{equation*}
S_n(y_1,y_2,\dots) = n(y_1+T) \, S_{n-1}(y_1,y_2,\dots) \,.
\end{equation*}
Due to \eqref{e3.10}, the latter is equivalent to the obvious identity
\begin{equation*}
(y_1+T) \, \exp \Bigl( \sum_{k=1}^\infty z^k y_k \Bigr)
= \d_z \, \exp \Bigl( \sum_{k=1}^\infty z^k y_k \Bigr) \,.
\end{equation*}
\end{remark}

Equation \eqref{e-apr12-3} (or \eqref{e3.13})
gives the most general $\la$-bracket of $e$ and $f$
that satisfies the Jacobi identity of type $hef$ for $\alpha\neq0$. 
By rescaling the generator $e$, we can assume that the constant
$\delta=1$.
Next, we are going to impose the Jacobi identity of type $eef$. 
In the vertex algebra case, we will do this separately for $\al\ne0$
and $\al=0$ in Sections \ref{s-eef1} and \ref{s-eef0}, respectively.
The Poisson case will be discussed in Section \ref{s-eefP}.

\section{Jacobi identity for $e,e,f$ with $\alpha\neq0$}
\label{s-eef1}

In this section, we will work in a vertex algebra solving Problem \ref{p1},
or a non-linear Lie conformal algebra solving Problem \ref{p2}.
We will assume that $\al\ne0$, and we let $\be=1/\al$. The results
of the previous section determine the $\la$-bracket $[e_\la f]$, i.e.,
the polynomial $P$ from \eqref{e1.1}.
We are going to impose the Jacobi identity of type $eef$:
\begin{equation}\label{e5.1}
[e_\la [e_\mu f]] = (-1)^{p(e)} \, [e_\mu [e_\la f]] \,,
\end{equation}
which will determine the possible values of $\al$.

It follows from Proposition \ref{p-apr12-1} that 
equation \eqref{e5.1} can be deduced from the $\la$-brackets \eqref{e1.1}
and the axioms of vertex algebra
if and only if we have a non-linear Lie conformal algebra 
$R=\CC[T]\langle h,e,f\rangle$
satisfying the assumptions of Problem \ref{p2}.
More generally, equation \eqref{e5.1} may hold under some additional relations among 
the generators $h,e,f$ and their derivatives;
in this case we may have a vertex algebra satisfying the assumptions of Problem \ref{p1},
which is not freely generated.
In conclusion, in order to solve Problems \ref{p1} and \ref{p2} (namely, in order to prove
Theorems \ref{th-june12} and \ref{th-june12b}) we need to study equation \eqref{e5.1}.

\subsection{Proof of Proposition \ref{pnlie}}\label{spfal-1}

We are given the $\lambda$-bracket \eqref{e1.1} and
\begin{equation*}
[e_\la f] = {:}P(\la;h,Th,\dots){:} = {:}p(\lambda+T-h) 1 {:} \,,
\end{equation*}
where the generators $h,e,f$ are all even, $\al=-1$ and $p\in\CC[\la]$
is an arbitrary polynomial. We need to check the Jacobi identities
for the triples $(h,e,f)$, $(e,e,f)$, and $(f,f,e)$.

As before, let $\be=1/\al=-1$.
The Jacobi identity for $(h,e,f)$  follows immediately 
from Proposition \ref{p-apr12-1}. Next, applying Corollary \ref{clabe} 
for the elements $a=e$, $b=-h$, we obtain:
\begin{equation*}
[e_\la [e_\mu f]] = {:} p'(\la+\mu+T-h) \, e {:} \,,
\end{equation*}
which makes the Jacobi identity of type $eef$ obvious (see \eqref{e5.1}).
The one of type $ffe$ can be checked in a similar way, or, alternatively,
derived from symmetry considerations. Indeed, by the remarks in
Section \ref{sgrsym}, we can assume that $p(\la)=\la^d$ is homogeneous.
Then Lemma \ref{l-july17-1} implies $\tilde P = (-1)^{d+1} P$, and
the Jacobi identity of type $ffe$ follows from the observations in Section \ref{sgrsym}.
Therefore, $R=\CC[T]\langle h,e,f\rangle$ satisfies
all axioms of a non-linear Lie conformal algebra.
Recall that when $p(\la)=\la^d$ this algebra is denoted as $R_{-1}^d$.
\qed

\subsection{Computation of the commutator}\label{scomcom}

As before, we assume that the polynomial $P$ is homogeneous of degree $d\ge1$
with respect to the grading given by $\deg\la=\deg T=\deg h=1$.
Then $P$ is determined up to a constant by Proposition \ref{p-apr12-1}(b).
By rescaling the generator $e$, we can take the constant
$\delta=1/d!$ in \eqref{e-apr12-3} (corresponding to $\gamma=1$ in 
\eqref{e3.13}), so that
\begin{equation}\label{june134}
[e_\la f] = {:} P(\la;h,Th,\dots) {:} 
= \Res_z \frac{e^{z\lambda}}{z^{d+1}} \; {:} e^{z(T+\beta h)} 1 {:} \,.
\end{equation}

To find the double commutator $[e_\la [e_\mu f]]$, we apply 
Proposition \ref{pro-june12} for the elements $a=e$, $b=\be h$,
and we obtain from equation \eqref{june12-d} that
\begin{equation*}
\bigl[ e_\la {:} e^{z(T+\beta h)} 1 {:} \bigr] 
= {:} (e^{z(T+\beta h)} 1) \, \big(\Psi_\be(z,\lambda+T) e\big) {:} \,,
\end{equation*}
where the function $\Psi_\be$ is defined by \eqref{fbe3}. Therefore,
\begin{equation}\label{june136}
[e_\la [e_\mu f]] = \Res_z \frac{e^{z\mu}}{z^{d+1}} \;
{:} (e^{z(T+\beta h)} 1) \, \big(\Psi_\be(z,\lambda+T) e\big) {:} \,.
\end{equation}
Now we will compute the coefficients of certain powers of $\la$ and $\mu$
in this formula.

\begin{lemma}\label{lem-july18}
The right-hand side of\/ \eqref{june136} has the following expansion
as a polynomial of\/ $\la,\mu${\rm:}
\begin{align*}
\sum_{k=0}^{d-1} \, \frac{\la^{k}}{k!} & \frac{\mu^{d-1-k}}{(d-1-k)!} \,
\binom{\be}{k+1} (-1)^{k+1} e
\\
&+ \frac{\la^{d-2}}{(d-2)!} \, 
\Bigl\{ \binom{\be}{d-1} (-1)^{d-1} \be \, {:}he{:}
+ \binom{\be}{d} (-1)^{d} \, Te \Bigr\} 
\\
&+ \frac{\mu^{d-2}}{(d-2)!} \,
\Bigl\{ -\be^2 \, {:}he{:} + \binom{\be}{2} \, Te \Bigr\} 
+ \cdots \,,
\end{align*}
where the dots denote other powers of\/ $\la,\mu$.
\end{lemma}
\begin{proof}
{}From \eqref{fbe3}, we obtain the expansion
\begin{equation}\label{exp18}
\frac{e^{z\mu}}{z^{d+1}} \, \Psi_\be(z,\lambda+T)
= \sum_{l=0}^\infty \sum_{n=1}^\infty \, z^{l+n-d-1} \,
\binom{\be}n (-1)^n \frac{\mu^{l}}{l!} \frac{(\lambda+T)^{n-1}}{(n-1)!} \,.
\end{equation}
In order to get the coefficient of $\la^{k} \mu^{d-1-k}$ in \eqref{exp18}, we need
$l=d-1-k$ and $n-1 \ge k$. Then the power of $z$ is
$n-k-2 \ge -1$. Therefore, under the residue in $z$
we must have $n-1=k$ and
the exponential $e^{z(T+\beta h)}$ in \eqref{june136} can be replaced by 1.
Hence, the coefficient in front of 
$\frac{\la^{k}}{k!} \frac{\mu^{d-1-k}}{(d-1-k)!}$ in \eqref{june136} is exactly
$\binom{\be}{k+1} (-1)^{k+1} e$.

Similarly, to get the coefficient of $\la^{d-2}$ in \eqref{exp18}, 
we need $l=0$ and $n-1 \ge d-2$. Then the power of $z$ is $n-d-1 \ge -2$. 
Hence, only the terms with $n=d-1$ and $n=d$ in \eqref{exp18} 
contribute to the residue with respect to $z$.
We thus obtain that the coefficient in front of $\frac{\la^{d-2}}{(d-2)!}$ in \eqref{june136} is
\begin{equation*}
\binom{\be}{d-1} (-1)^{d-1} \, {:} ((T+\beta h) 1) \, e {:}
+ \binom{\be}{d} (-1)^{d} \, Te \,,
\end{equation*}
as claimed.

Finally, in order to have the coefficient of $\mu^{d-2}$ in \eqref{exp18}, 
we need $l=d-2$. Then the power of $z$ is $n-3 \ge -2$.
Hence, the only terms in \eqref{exp18} contributing to the residue in $z$ are
for $n=1$ and $n=2$.
It follows that the coefficient of $\frac{\mu^{d-2}}{(d-2)!}$ in \eqref{june136} is
$$
-\beta{:}\big((T+\beta h)1\big)e{:}+\binom{\beta}{2}Te\,,
$$
thus completing the proof.
\end{proof}

\subsection{Proof of Theorems \ref{th-june12} and \ref{th-june12b} for 
$\al\ne0$}\label{spfalnz}

First of all, notice that by Proposition \ref{pnlie},
$R_{-1}^d$ is indeed a non-linear Lie conformal algebra solving 
Problem \ref{p2}. Then $V^d_{-1}:=V(R_{-1}^d)$ is a vertex algebra solving 
Problem \ref{p1}. By the results of Section \ref{svbe}, 
the lattice vertex algebra $V_{\ZZ\sqrt{d+1}}$ also solves Problem \ref{p1}.

Now consider a vertex algebra solving Problem \ref{p1},
or a non-linear Lie conformal algebra solving Problem \ref{p2},
with a homogeneous polynomial $P$ of degree $d\geq1$.
As discussed at the beginning of Section \ref{scomcom},
the $\la$-bracket $[e_\la f]$ is given by \eqref{june134},
and this determines the polynomial $P$. To find the possible values
of $\al$, we will impose the Jacobi identity \eqref{e5.1}, where
$[e_\la [e_\mu f]]$ is given by \eqref{june136}.
Let us compare the coefficients of $\frac{\la^{k}}{k!} \frac{\mu^{d-1-k}}{(d-1-k)!}$
in both sides of \eqref{e5.1} for $k=0,1,\dots,d-1$.
Then Lemma \ref{lem-july18}
leads to the following system of equations:
\begin{equation}\label{andrea2}
\binom{\beta}{k+1} (-1)^{k+1}
= (-1)^{p(e)} \binom{\beta}{d-k} (-1)^{d-k}
\,, \qquad k=0,1,\dots,d-1 \,.
\end{equation}

For $d=1$, we obtain a single equation, $\be=(-1)^{p(e)} \be$,
which, since $\be\ne0$, gives $p(e)=\bar0$ and is then satisfied
for every $\be$. In this case, by the symmetry \eqref{symef}, 
we also have $p(f)=\bar0$.
After rescaling, this produces the current Lie conformal algebra
or the corresponding vertex algebra 
(see parts (a) in Theorems \ref{th-june12} and \ref{th-june12b}).

The case $d\geq2$ is treated in the following lemma.

\begin{lemma}\label{lem-june12}
Given an integer $d\geq2$ and $p(e)\in\ZZ/2\ZZ$, 
all nonzero complex numbers $\be$
satisfying equations \eqref{andrea2} are{\rm:}
\begin{enumerate}
\item $\be=-1$, if\/ $p(e)=\bar0$.
\item $\be=d+1$, if\/ $p(e)=\overline{d+1}$.
\end{enumerate}
\end{lemma}
\begin{proof}
It is easy to check that (a) and (b) are indeed solutions of equations
\eqref{andrea2}. Now assume that $\be\ne0$ is a solution.
Our first simple observation is that $\be\not\in\{1,\dots,d-1\}$.
For otherwise, taking $k=0$ in \eqref{andrea2} would give $\be=0$.
Next, we will consider separately the cases when $d$ is odd or even.

Assume first that $d=2n+1$ is odd. Then \eqref{andrea2} with $k=n$ 
implies $p(e)=\bar0$. Next, \eqref{andrea2} with $k=n-1$ gives 
$\binom{\be}{n} = \binom{\be}{n+2}$, which is equivalent to
\begin{equation*}
\frac{(\be-n)(\be-n-1)}{(n+1)(n+2)} = 1\,.
\end{equation*}
The latter has only two solutions: $\be=-1$ and $\be=2n+2=d+1$.

Now if $d=2n$ is even, equation \eqref{andrea2} for $k=n$ gives
$\binom{\be}{n+1} = \binom{\be}{n} (-1)^{p(e)+1}$, which is equivalent to
\begin{equation*}
\frac{\be-n}{n+1} = (-1)^{p(e)+1}\,.
\end{equation*}
Therefore, either 
$p(e)=\bar0$ and $\be=-1$, or $p(e)=\bar1$ and $\be=2n+1=d+1$.
\end{proof}

Continuing with the proof of Theorems \ref{th-june12} and \ref{th-june12b},
we need to consider the two cases from Lemma \ref{lem-june12}.
When $\be=-1$ and $p(e)=\bar0$, we have $\alpha=-1$,  
and by the symmetry \eqref{symef}, $p(f)=\bar0$. 
This gives the $\la$-brackets of the 
non-linear Lie conformal algebra $R_{-1}^d$.
The associated enveloping vertex algebra $V^d_{-1}=V(R_{-1}^d)$
is freely generated. By the results of \cite{DSK1}, any other vertex algebra 
satisfying the same $\la$-bracket relations among the generators
is a quotient of $V^d_{-1}$.
In this way, we obtain parts (b) in 
Theorems \ref{th-june12} and \ref{th-june12b}.

When $\be=d+1$ and $p(e)=\overline{d+1}$, we compare the
coefficients of $\la^{(d-2)}$ in both sides of \eqref{e5.1},
using again Lemma \ref{lem-july18}. We obtain the equation
\begin{eqnarray*}
&& (d+1)\binom{d+1}{d-1} (-1)^{d-1} \, {:}he{:}
+ \binom{d+1}{d} (-1)^{d} \, Te \\
&& \,\,\, \,\,\, = (-1)^{d+1} \, \Bigl( -(d+1)^2 \, {:}he{:} + \binom{d+1}{2} \, Te \Bigr)
\,,
\end{eqnarray*}
which simplifies to
$Te = (d+1) \, {:}he{:}\,$.
This is exactly the first relation in \eqref{vbe5}, and the
second one can be deduced by symmetry considerations
(see Section \ref{sgrsym}).
Because \eqref{vbe5} are additional relations, not part of the vertex algebra axioms,
it follows that non-linear Lie conformal algebras solving
Problem \ref{p2} with $\alpha=(d+1)^{-1}$ do \emph{not} exist.
This completes the proof of Theorem \ref{th-june12} in the case when
$\al\ne0$.

We showed that, in any vertex algebra solving Problem \ref{p1} 
with $\alpha=(d+1)^{-1}$, the generators satisfy relations \eqref{vbe5}. 
In addition, by the results of Section \ref{ssoleq}, 
the polynomial $P$ is given by \eqref{vbe3}.
It is known that there exists a unique (up to isomorphism)
such a vertex algebra, namely the lattice vertex algebra $V_{\ZZ\sqrt{d+1}}$
(see e.g.\ Sections 5.4 and 5.5 in \cite{K}).
This completes the proof of Theorem \ref{th-june12b}.

\begin{remark}\label{rem-june12}
{}From the above proof we can deduce that, in addition of being
not freely generated, the lattice vertex algebra $V_{\ZZ\sqrt{d+1}}$ 
cannot be realized as a quotient of a freely generated vertex algebra by 
an ``irregular'' ideal (see \cite{DSK1} for the definition). 
\end{remark}

To complete the proof of Theorem \ref{th-june12}, we have to consider the case
$\al=0$. This will be done in the next section.

\section{Jacobi identity for $e,e,f$ with $\alpha=0$}
\label{s-eef0}

Now we will consider a non-linear Lie conformal algebra $R$ 
solving Problem \ref{p2} with $\al=0$. 
We denote by $V=V(R)$ its universal enveloping  vertex algebra, 
which is freely generated by the elements $h,e,f$.

\subsection{Differential polynomials in $h$}\label{s-dh}

Consider the polynomial ring $\CC[h,Th,\dots]$,
equipped with the derivation $T$, 
and the subalgebra $\langle h\rangle\subset V$ generated by $h$.

\begin{lemma}\label{lem-dh}
In any non-linear Lie conformal algebra solving Problem \ref{p2} with $\al=0$,
we have{\rm:}
\begin{enumerate}
\item 
The $\la$-bracket of any two elements of\/ $\langle h\rangle$ is zero.

\item
The normally ordered product in $\langle h\rangle$ 
is commutative and associative.

\item
The map $\CC[h,Th,\dots]\to\langle h\rangle$
defined by \eqref{inj} is an associative algebra isomorphism.
\end{enumerate}
\end{lemma}
\begin{proof}
First of all, $[h_\la h]=0$ implies $[h_\la (T^k h)]=0$ for all $k\in\ZZ_+\,$.
{}From the Wick formula \eqref{wick} we deduce by induction that
$[h_\la {:} (T^{k_1} h)\cdots(T^{k_s} h) {:}]=0$
for arbitrary $k_1,\dots,k_s\in\ZZ_+$.
Then by the skew-symmetry \eqref{ss}, we deduce that
$[a_\la b]=0$ for all $a,b\in\langle h\rangle$. 
This proves part (a).
Part (b) follows from (a), the quasi-commutativity \eqref{qcom}, and 
quasi-associativity \eqref{qas}.
Then (c) is immediate from (b) and the fact that $\langle h\rangle$
is freely generated.
\end{proof}

We introduce the elements
\begin{equation*}
H^k=(T+h)^k\,1 \in \CC[h,Th,T^2h,\dots]\,, \qquad k\in\ZZ_+ \,,
\end{equation*}
where on the right we use that $T1=0$. For example,
\begin{equation*}
H^0=1 \,, \qquad H^{1}=h \,, \qquad 
H^{2}=Th+h^2 \,, \qquad H^3=T^2h+3h(Th)+h^3 \,.
\end{equation*}
Notice that $H^{k+1}=T^k h$ $+$ 
terms involving lower order derivatives of $h$.
Thus, we have an isomorphism $\CC[h,Th,\dots]\simeq\CC[H^1,H^2,\dots]$.
As usual, the image of $H^k$ in $\langle h\rangle$ under the map \eqref{inj}
will be denoted as ${:}H^k{:}={:}(T+h)^k \, 1{:}$. In particular,
${:}H^0{:}={:}1{:}=\vac$.
Similar remarks apply to the elements $\tilde H^k=(T-h)^k\,1$.
Notice that, up to a sign, $\tilde H^k$ is the image of $H^k$
under the change of variables \eqref{symef}.

Applying Corollary \ref{clabe} for $\be=1$ and the elements $a=e$, $b=h$,
we obtain:
\begin{equation}\label{eqb}
[e_\la {:}H^k{:}] = -k \, {:}H^{k-1}e{:} \,, \qquad k\in\ZZ_+ \,.
\end{equation}
Here we take, as usual, normally ordered products of more than two elements
from right to left. 
Similarly, Corollary \ref{clabe} for $\be=1$ and $a=f$, $b=-h$ gives
\begin{equation*}
[f_\la {:}\tilde H^k{:}] = -k \, {:}\tilde H^{k-1}f{:} 
\,,  \qquad k\in\ZZ_+ \,.
\end{equation*}
This equation can also be deduced from \eqref{eqb} by applying
the symmetry \eqref{symef}.

For any polynomial $v(\lambda)=\sum_{n=0}^N v_n\lambda^n \in V[\lambda]$
with $v_N \ne0$, we will denote by $\lt v(\lambda) := v_N \lambda^N$ 
its \emph{leading term}.

\begin{lemma}\label{l-july20}
For every $A\in\CC[h,Th,\dots]$, $b\in\langle h\rangle$,
and\/ $k_1,\dots,k_s\in\NN$, we have{\rm:}
\begin{enumerate}
\item 
$\lt [{:}Ae{:}_\lambda b] = {:}A (\lt [e_\lambda b]) {:} \,,$
\medskip

\item 
$\lt [e_\la{:}H^{k_1}\cdots H^{k_s}{:}]
= \dfrac{\lambda^{s-1}}{(s-1)!} \, (-1)^s k_1\cdots k_s \, 
{:}H^{k_1-1}\cdots H^{k_s-1}e{:} \,.$
\end{enumerate}
\end{lemma}
\begin{proof}
Part (a) is trivial for $A=1$. By induction, we are going to prove it for 
monomials $A=(T^{k_1} h)\cdots(T^{k_s} h)$. Write
\begin{equation*}
A=(T^k h)C \,, \qquad\text{where}\quad
k=k_1 \,, \; C = (T^{k_2} h)\cdots(T^{k_s} h) \,,
\end{equation*}
and $C=1$ when $s=1$. 
Then ${:}Ae{:} = {:}(T^k h)({:}Ce{:}){:}$ by the convention
of taking normally ordered product from right to left (cf.\ \eqref{mono2}).
Using the right Wick formula \eqref{rwick} and Lemma \ref{lem-dh}(a), we obtain
\begin{equation*}
[{:}Ae{:}_\lambda b]  
= [{:}(T^k h)({:}Ce{:}){:}_\la b]
= {:} (e^{T \d_\la} T^k h) [{:}Ce{:}_\la b] {:} \,.
\end{equation*}
Taking the leading terms and applying the inductive assumption, we get
\begin{equation*}
\lt [{:}Ae{:}_\lambda b] 
= {:} (T^k h) \bigl( \lt [{:}Ce{:}_\la b] \bigr) {:} 
= {:} (T^k h) \bigl( {:}C (\lt [e_\la b]) {:} \bigr) {:} 
= {:}A (\lt [e_\lambda b]) \,,
\end{equation*}
which proves part (a).

For $s=1$, part (b) follows immediately from \eqref{eqb}.
For $s\ge2$, we proceed by induction on $s$. 
Due to Lemma \ref{lem-dh}(b), we can write
\begin{equation*}
a = {:}H^{k_1}\cdots H^{k_s}{:} = {:}H^{k_1}b{:} = {:}({:}H^{k_1}{:})b{:} 
\,, \qquad
b = {:}H^{k_2}\cdots H^{k_s}{:} \,.
\end{equation*}
Then applying the noncommutative Wick formula \eqref{wick}, we obtain
\begin{equation*}
[e_\la a] 
= {:}[e_\lambda {:}H^{k_1}{:}]b{:}
+ {:}({:}H^{k_1}{:})[e_\lambda b]{:}
+ \int_0^\lambda d\mu \, [[e_\lambda{:}H^{k_1}{:}]_\mu b] \,.
\end{equation*}
The first two terms in the right-hand side have degree in $\la$ at most $s-2$,
by \eqref{eqb} and the inductive assumption. Thus,
\begin{equation*}
\lt [e_\la a] 
= \lt \int_0^\lambda d\mu \, [[e_\lambda{:}H^{k_1}{:}]_\mu b]
= -k_1 \int_0^\lambda d\mu\, \lt [{:}H^{k_1-1}e{:}_\mu b]{:}\,,
\end{equation*}
where we again used \eqref{eqb}.
Now part (a) and the inductive assumption imply
\begin{align*}
\lt [{:}H^{k_1-1} & e{:}_\mu b]{:}
= {:}H^{k_1-1} \bigl( \lt [e_\mu b] \bigr){:}
\\
&= \frac{\mu^{s-2}}{(s-2)!} \, (-1)^{s-1} k_2\cdots k_s \, 
{:}H^{k_1-1} H^{k_2-1}\cdots H^{k_s-1}e{:} \,,
\end{align*}
which completes the proof.
\end{proof}

The same proof as above (or the symmetry \eqref{symef}) gives:
\begin{equation}\label{eqb4}
\lt [f_\la{:}\tilde H^{k_1}\cdots \tilde H^{k_s}{:}]
= \frac{\lambda^{s-1}}{(s-1)!} \, (-1)^s k_1\cdots k_s \, 
{:}\tilde H^{k_1-1}\cdots \tilde H^{k_s-1}f{:} \,.
\end{equation}

\subsection{Proof of Theorem \ref{th-june12} for $\al=0$}\label{spfalz}

Consider a non-linear Lie conformal algebra solving Problem \ref{p2} 
with $\al=0$. 
By Proposition \ref{p-apr12-1}, the Jacobi identity for elements $h,e,f$
holds if and only if 
\begin{equation*}
[e_\lambda f] = {:}P(h,Th,T^2 h,\dots){:}
\end{equation*}
is independent of $\la$.
As before, we assume that the polynomial $P$ is homogeneous of degree $d\ge1$
with respect to the grading given by $\deg T=\deg h=1$.

Then the Jacobi identity
\begin{equation*}
[e_\lambda [e_\mu f]] = (-1)^{p(e)} \, [e_\mu [e_\lambda f]]
\end{equation*}
implies that the polynomial 
$[e_\lambda {:}P(h,Th,\dots){:}]$
does not depend on $\la$. Using Lemma \ref{l-july20}(b), 
it is easy to see that this is possible only when
$P(h,Th,\dots) = \gamma H^d$ for some $\gamma\in\CC$.
After rescaling the generator $e$, we can take $\gamma=1$.

On the other hand, by the skew-symmetry \eqref{ss}, we have
\begin{equation*}
[f_\lambda e] = -(-1)^{p(e)p(f)} \, {:}P(h,Th,\dots){:} \,.
\end{equation*}
In the same way as above, using \eqref{eqb4}, the Jacobi identity 
\begin{equation*}
[f_\lambda [f_\mu e]] = (-1)^{p(f)} \, [f_\mu [f_\lambda e]]
\end{equation*}
implies that $[f_\lambda e]$ is a scalar multiple of $\tilde H^d$.
Therefore,
$P(h,Th,\dots) = H^d = \gamma \tilde H^d$ for some $\gamma\in\CC$,
which is possible only for $d=1$. 
We thus obtain the current Lie conformal algebra $\CC[T]\sl_2$ at level $0$,
thus completing the proof of Theorem \ref{th-june12}.
\qed

\begin{remark}\label{r-pfalz}
The above proof also works for a vertex algebra $V$ solving Problem \ref{p1}
with $\al=0$, provided the following additional assumptions hold:
\begin{enumerate}
\item 
The subalgebra $\langle h\rangle\subset V$ generated by $h$ 
is freely generated, i.e., the map $\CC[h,Th,\dots]\to V$
defined by \eqref{inj} is injective.

\item 
The similarly defined map 
$P(h,Th,\dots)$ $\mapsto$ ${:}P(h,Th,\dots) e{:}$
is injective.
\end{enumerate}
The conclusion is then that $V$ is a quotient 
of the universal enveloping  vertex algebra of the current Lie 
conformal algebra $\CC[T]\sl_2$ at level $0$.
\end{remark}

\section{The Poisson case}\label{s-eefP}

Consider a Poisson vertex algebra $\mc V$ solving Problem \ref{p1},
or a non-linear Poisson conformal algebra $R$ 
solving Problem \ref{p2}. In the latter case, we denote
by $\mc V=\Sym(R)$ its universal enveloping Poisson vertex algebra.
As before, we assume that the polynomial 
$P(\la;x_1,x_2,\dots)$
from \eqref{e1.1}
is homogeneous of degree $d\ge1$
with respect to the grading given by $\deg\la=1$, $\deg x_k=k$,
where $x_k = T^{k-1} h$.

In this section, we are going to impose the Jacobi identity of type $eef$
(see \eqref{e5.1}). 
Using the Leibniz rule \eqref{xxx1}, we have:
\begin{equation}\label{eefp1}
\begin{split}
[e_\la [e_\mu f]] &= [e_\la P(\mu;h,Th,\dots)]
\\
&= \sum_{k=1}^\infty \, [e_\la (T^{k-1} h)] \;
\frac{\d P}{\d x_k}(\mu;h,Th,\dots)
\\
&= -\sum_{k=1}^\infty \, 
\frac{\d P}{\d x_k}(\mu;h,Th,\dots) \, 
\bigl( (\la+T)^{k-1} e \bigr) \,.
\end{split}
\end{equation}

\subsection{Proof of Theorem \ref{th-pc1} for $\al=0$}

By Proposition \ref{p-apr12-1}, for $\al=0$, the Jacobi identity 
of type $hef$ holds if and only if 
$P(\la;h,Th,\dots)=P(h,Th,\dots)$
is independent of $\la$. Due to \eqref{eefp1},
equation \eqref{e5.1} can be written as
\begin{equation*}
\sum_{k=1}^\infty \, 
\frac{\d P}{\d x_k}(h,Th,\dots) \, 
\bigl( (\la+T)^{k-1} e \bigr)
= (-1)^{p(e)} \, \sum_{k=1}^\infty \, 
\frac{\d P}{\d x_k}(h,Th,\dots) \, 
\bigl( (\mu+T)^{k-1} e \bigr) \,.
\end{equation*}
Since now we assume $\mc V$ is freely generated, 
the above equation is equivalent to:
\begin{equation*}
p(e)=\bar0 \quad\text{and}\quad
\frac{\d P}{\d x_k} = 0 \,, \quad k\ge2 \,.
\end{equation*}
Therefore, $P(h,Th,\dots)=p(h)$ is a polynomial of $h$.
(When $P$ is homogeneous, after rescaling the generator $e$, 
we can assume that $P=h^d$.)
Similarly, the Jacobi identity of type $ffe$ gives that $p(f)=\bar0$.
This proves Theorem \ref{th-pc1} in the case $\al=0$.
\qed

\begin{remark}\label{r-ppalz}
Let $\mc V$ be a Poisson vertex algebra solving Problem \ref{p1}
with $\al=0$. Then the above proof works provided the following additional 
assumptions hold:
\begin{enumerate}
\item 
The subalgebra $\langle h\rangle\subset \mc V$ generated by $h$ 
is freely generated, i.e., $\langle h\rangle$ is isomorphic 
to $\CC[h,Th,\dots]$.

\item 
The map $\langle h\rangle\to V$ given by
$b \mapsto be$ is injective.
\end{enumerate}
The conclusion is then that $\mc V$ is a quotient of $\Sym(R)$, where
$R$ is one of the non-linear Poisson conformal algebras described in 
Theorem \ref{th-pc1}.
\end{remark}

\subsection{Proof of Theorems \ref{th-pc1} and \ref{t4.1} for $\al\ne0$}

Let now $\al\ne0$ and $\be=1/\al$. By Lemma \ref{l-july16}, when 
the polynomial $P$ is homogeneous of degree $d\ge1$,
it is given explicitly by formula \eqref{e3.9}.
Then \eqref{eefp1} can be rewritten as
\begin{equation}\label{eefp4}
[e_\la [e_\mu f]] = - \gamma\, \sum_{k=1}^\infty \, 
\frac{\d}{\d x_k} \, S_d
\Bigl(\mu+\be\frac{x_1}{1!} \,,\,
\be\frac{x_2}{2!} \,,\, \be\frac{x_3}{3!} \,,\, \dots\Bigr) 
\bigl( (\la+T)^{k-1} e \bigr) \,,
\end{equation}
where $x_k = T^{k-1} h$.
Using the definition \eqref{vbe4} of $S_d$, it is easy to see that
the coefficient of $\la^{d-1}$ in the right-hand side of \eqref{eefp4} is 
$-\beta\gamma e / d!$, while the coefficient of $\mu^{d-1}$ is
$-\beta\gamma e / (d-1)!$.

Hence, the Jacobi identity \eqref{e5.1} implies
$d=(-1)^{p(e)}$, which is possible only when $d=1$ and $e$ is even.
By symmetry, $f$ must also be even. In this case, equations
\eqref{e1.1} correspond to the $\la$-brackets of
the current Lie conformal algebra $R=\CC[T]\sl_2$.
This completes the proof of Theorems \ref{th-pc1} and \ref{t4.1}.
\qed

\section{Wakimoto realization of $V^d_{-1}$}\label{s-wak}

In this section, we describe a free-field realization of (a quotient of) 
the vertex algebra $V^d_{-1}$ defined in Theorem \ref{th-june12b}(b).
This gives a representation of $V^d_{-1}$ on the Fock space, 
which generalizes the Wakimoto realization
of the affine vertex algebra of $\sl_2$ at the critical level $-2$
(see \cite{W}).

\subsection{Computations in the Fock space}
Consider the non-linear Lie conformal algebra
$R=\CC[T]\langle a,b \rangle$ with two even generators $a,b$
and $\la$-brackets
\begin{equation}\label{e8.1}
[a_\la a]=[b_\la b]=0\,,\qquad [a_\la b]=-[b_\la a]=\vac\,.
\end{equation}
The corresponding universal enveloping vertex algebra $\mc F=V(R)$
is called the \emph{Fock space}. The quantum fields $Y(a,z)$ and
$Y(b,z)$ corresponding to the elements $a,b\in\mc F$ are 
known as \emph{free bosons} (cf.\ \cite[Section 3.5]{K}). 
By formula \eqref{comfla}, 
they satisfy the commutation relations:
\begin{equation*}
[Y(a,z),Y(a,w)] = [Y(b,z),Y(b,w)] = 0 \,, \qquad
[Y(a,z),Y(b,w)] = \delta(z-w) \,.
\end{equation*}
%
We introduce the following elements of $\mc F$:
\begin{equation}\label{e-wak1}
H=-{:}ab{:} \,,\quad
E_n={:}(H+T)^na{:} \,,\quad
F_n={:}(H-T)^nb{:} \,,\qquad n\in\ZZ_+ \,.
\end{equation}

\begin{lemma}\label{l-wak1}
For every $m,n\geq0$, the above elements satisfy{\rm:}
\begin{align}
\label{e8.3}
[H_\la a]&=-[a_\la H]=a \,,\qquad [H_\la b]=-[b_\la H]=-b\,,
\\ \label{e8.4}
[H_\la H]&=-\la\vac \,, \qquad 
[H_\la E_n]=E_n \,,\qquad 
[H_\la F_n]=-F_n \,,
\\ \label{e8.5}
[{E_n}_\la E_n]&=[{F_n}_\la F_n]=0 \,, \qquad
\d_\la [{E_m}_\la E_n]=\d_\la [{F_m}_\la F_n]=0 \,,
\\ \label{e8.6}
[{E_m}_\la F_n]&=(m+n+1) \, {:}(H-T-\la)^{m+n} \vac{:} \,,
\\ \label{e8.7}
{:}E_m F_n{:} &= -{:}(H-T)^{m+n+1} \vac{:} \,.
\end{align}
\end{lemma}
\begin{proof}
Equations \eqref{e8.3} follow easily from \eqref{e8.1}, \eqref{e-wak1},
the noncommutative Wick formula \eqref{wick}, and skew-symmetry \eqref{ss}.
Indeed, we have, for example,
\begin{equation*}
-[a_\la H] = [a_\la {:}ab{:}]
= {:}[a_\la a]b{:} + {:}a[a_\la b]{:} + \int_0^\la d\mu\, [[a_\la a]_\mu b]
= a \,.
\end{equation*}
Similarly,
\begin{equation*}
\begin{split}
-[H_\la H] &= [H_\la {:}ab{:}]
= {:}[H_\la a]b{:} + {:}a[H_\la b]{:} + \int_0^\la d\mu\, [[H_\la a]_\mu b]
\\
&= {:}ab{:} - {:}ab{:} + \la\vac = \la\vac \,,
\end{split}
\end{equation*}
thus proving the first equation in \eqref{e8.4}.
The second one, $[H_\la E_n]=E_n$, for $n=0$ becomes $[H_\la a]=a$.
We will prove it for all $n\ge0$ by induction.
Using that by definition
\begin{equation}\label{e-wak2}
E_{n+1} = {:}HE_n{:}+TE_n\,,\qquad 
F_{n+1} = {:}HF_n{:}-TF_n\,,
\end{equation}
we obtain from the noncommutative Wick formula, 
sesqui-linearity \eqref{ca}, and inductive assumption,
\begin{equation*}
[H_\la E_{n+1}] = [H_\la {:}HE_n{:}]+[H_\la (TE_n)]
= -\la E_n + {:}HE_n{:} +(\la+T)E_n = E_{n+1}\,.
\end{equation*}
The same argument can be used with $F_{n+1}$ instead of $E_{n+1}$.
This proves \eqref{e8.4}.

Next, we will prove that $\d_\la [{E_m}_\la E_n]=0$ by induction on $m+n$.
The case $m=n=0$ is trivial. Assume that $[{E_m} _\la E_n]$ is independent of 
$\la$, namely $[{E_m} _\la E_n]={E_m}_{(0)}{E_n}$. 
Then, using \eqref{e-wak2} and \eqref{e8.4}, we find
\begin{equation*}
\begin{split}
[{E_m} _\la {E_{n+1}}]
&= [{E_m} _\la {:}HE_n{:}]+[{E_m} _\la (TE_n)] 
\\
&= -{:}E_m E_n{:}
+ {:}H[{E_m}_\la E_n]{:}
- \int_0^\la d\mu \, [{E_m}_\mu E_n]
+ (\la+T) [{E_m}_\la E_n]
\\
&= -{:}E_m E_n{:}
+ {:}(H+T)({E_m}_{(0)}{E_n}){:} \,,
\end{split}
\end{equation*}
which is constant in $\la$.
By skew-symmetry,
\begin{equation*}
[{E_{n+1}}_\la E_m] = -[{(E_m)}_{-\la-T} E_{n+1}] 
= -[{E_m} _\la {E_{n+1}}]
\end{equation*}
is also independent of $\la$.
The above equation for $m=n+1$ implies $[{E_m}_\la E_m] = 0$,
thus completing the proof of \eqref{e8.5}.

We will prove \eqref{e8.6} and \eqref{e8.7} simultaneously,
again by induction on $m+n$. The case $m=n=0$ is obvious. 
Next, we consider the case $m\geq0$, $n=0$. 
By \eqref{e-wak2} and the quasi-associativity \eqref{qas}, we have
\begin{equation*}
\begin{split}
{:}E & {}_{m+1} F_0{:}
= {:}({:}HE_m{:})F_0{:} + {:}(TE_m)F_0{:}
\\
&= {:}H E_m F_0{:}
+ {:}\Bigl(\int_0^T d\la\, H\Bigr) [{E_m}_\la F_0]{:}
+ {:}\Bigl(\int_0^T d\la\, E_m\Bigr) [H_\la F_0]{:} + {:}(TE_m)F_0{:}
\,.
\end{split}
\end{equation*}
The third and fourth term in the right-hand side of this equation 
cancel each other.
By the inductive assumption, the first term is
\begin{equation*}
-{:}H(H-T)^{m+1} \vac{:} = - {:}H(H-T)^{m}H{:} \,,
\end{equation*}
while the second term is
\begin{equation*}
(m+1) \, {:}\Bigl(\int_0^T d\la\, H\Bigr) (H-T-\la)^{m} \vac{:}
= - {:}(H-T)^{m+1}H{:} + {:}H(H-T)^{m+1} \vac{:} \,,
\end{equation*}
because $T$ is a derivation.
We obtain equation \eqref{e8.7} with
$m+1$ in place of $m$ and $n=0$.
Similarly, using the right Wick formula \eqref{rwick}
and sesqui-linearity, we find
\begin{equation*}
\begin{split}
[& {E_{m+1}}_\la F_0]
= [{:}HE_m{:}_\la F_0] + [(T{E_m})_\la F_0]
\\ 
&= {:}(e^{T\d_\la}H) \, [{E_m}_\la F_0]{:}
+ {:}(e^{T\d_\la}E_m) \, [H _\la F_0]{:}
+\int_0^\la d\mu\, [{E_m}_\mu[H_\la F_0]]
- \la [{E_m}_\la F_0] \,.
\end{split}
\end{equation*}
By the inductive assumption and Taylor's formula,
the first term in the right-hand side is
\begin{equation*}
(m+1) \, {:}(e^{T\d_\la}H) (H-T-\la)^{m} \vac{:}
= (m+1) \, {:}(H-T-\la)^{m}H{:}\,.
\end{equation*}
It is easy to compute the other three terms and obtain \eqref{e8.6} with
$m+1$ in place of $m$ and $n=0$.
This proves equations \eqref{e8.6}, \eqref{e8.7} for $n=0$.

We are left with proving \eqref{e8.6} and \eqref{e8.7} 
for arbitrary $m,n \geq0$. 
Using \eqref{e-wak2}, \eqref{qcom2}, and \eqref{e8.4},
we get
\begin{equation*}
\begin{split}
{:}E_mF_{n+1}{:}
&= {:}E_m H F_n{:} - {:}E_m(TF_n){:}
\\
&= {:}H E_m F_n{:}  
+ {:} \Bigl( \int_{-T}^0 \, d\la \, [H_\la E_m] \Bigr) F_n {:}
-{:}E_m(TF_n){:} 
\\
&= {:}H E_m F_n{:} - {:}(TE_m)F_n{:} -{:}E_m(TF_n){:} \,.
\end{split}
\end{equation*}
Since $T$ is a derivation, this equals ${:}(H-T)({:}E_m F_n{:}){:}$,
as required.
Finally, to prove equation \eqref{e8.6}, we compute
\begin{equation*}
\begin{split}
[{E_m}_\la F_{n+1}]
&= [{E_m}_\la {:}HF_n{:}] - [{E_m}_\la (TF_n)]
\\
&= - {:}{E_m} F_n{:}
+ {:}H[{E_m}_\la F_n]{:} 
- \int_0^\la d\mu\, [{E_m}_\mu F_n] 
- (\la+T)[{E_m}_\la F_n] \,.
\end{split}
\end{equation*}
Using the inductive assumption, it is easy to derive from here
\eqref{e8.6} with $n+1$ instead of $n$.
This completes the proof of Lemma \ref{l-wak1}.
\end{proof}

\subsection{Wakimoto realization of $V^d_{-1}$}

Recall that the non-linear Lie conformal algebra $R_{-1}^d$ is defined
in Theorem \ref{th-june12}(b), and $V^d_{-1}:=V(R_{-1}^d)$ is its
universal enveloping  vertex algebra.

\begin{theorem}\label{t-wak1}
For every $d\geq1$ and every $n=0,\dots,d$, there is a vertex algebra
homomorphism $\pi_n\colon V^d_{-1}\to \mc F$,
given by
\begin{equation*}
\pi_n(h)=H\,,\qquad \pi_n(e)=E_n\,,\qquad 
\pi_n(f)=\frac{(-1)^d}{d+1} \, F_{d-n}\,.
\end{equation*}
\end{theorem}
\begin{proof}
Since $V^d_{-1}$ is freely generated, we only need to check that the 
$\la$-brackets among the generators $h,e,f$ are preserved. 
This is done by comparing 
equations \eqref{e1.1} with $\alpha=-1$ and \eqref{nlie5} to
equations \eqref{e8.4}--\eqref{e8.6} from Lemma \ref{l-wak1}.
\end{proof}

The map $\pi_n$ is not injective, because by \eqref{e8.7} we have
\begin{equation}\label{e8.9}
{:}(T-h)^{d+1} \vac{:} - (d+1) \, {:}ef{:}
\,\in\,\ker\pi_n
\end{equation}
for all $n=0,\dots,d$.
We define \emph{conformal weights} in $\mc F$ by 
$\Delta(a)=1$, $\Delta(b)=0$, 
and we obtain from \eqref{e-wak1} and \eqref{wtla} that
\begin{equation*}
\Delta(H)=1 \,, \qquad 
\Delta(E_n)=n+1 \,, \qquad
\Delta(F_n)=n \,.
\end{equation*}
Therefore, the map $\pi_n$ preserves the conformal weight if we let
$\Delta(h)=1$, $\Delta(e)=n+1$, and $\Delta(f)=d-n$ in $V^d_{-1}$
(cf.\ \eqref{e1.2}).

\begin{remark}\label{r-wak1}
The Wakimoto realization \cite{W} of the affine Lie algebra $\what{\sl}_2$ at
the critical level $-2$ is given by:
\begin{equation*}
\bar h = 2H = - 2 \, {:} ab {:}  \,, \qquad
e = E_0 = a \,, \qquad
f = F_1 = {:} (H-T)b {:} = - {:} ab^2 {:} - 2Tb
\end{equation*}
(in the last equality we used the quasi-associativity \eqref{qas}).
Indeed, it follows from Lemma \ref{l-wak1} that $\bar h,e,f$
satisfy equations \eqref{sl-1} with $k=-2$.
\end{remark}

\section{Zhu algebra of $V^d_{-1}$}\label{s-zhu}

In this section, we determine the Zhu algebra of the vertex algebra 
$V_{-1}^d$ with respect to the Hamiltonian operator defined by assigning 
conformal weights on the generators $h,e,f$ according to \eqref{e1.2}.
The result is one of the associative algebras introduced by Smith \cite{S}.

\subsection{Definition of the Zhu algebra}
Recall that a \emph{Hamiltonian operator} $H$ on a vertex algebra $V$ 
is a diagonalizable linear operator on $V$ such that the 
\emph{conformal weight} defined by $Ha=\Delta(a) \, a$ satisfies 
equations \eqref{wtla} for all $j\in\ZZ$
(see e.g.\ \cite[Section 4.9]{K} for more details).
We are also going to use the notation $\Delta_a$ for $\Delta(a)$.
We introduce the following $*_n$-products and  
$*$-bracket on $V$ (cf.\ \cite{Z,BK}):
\begin{align}
\label{e-zhu1a}
a*_nb &= \sum_{j\in\ZZ_+}\binom{\Delta_a}{j} \, a_{(n+j)}b \,,
\qquad\quad n\in\ZZ \,,
\\ \label{e-zhu1}
[a_* b] &= \sum_{j\in\ZZ_+}\binom{\Delta_a-1}{j} \, a_{(j)}b
= [a_{\partial_x}b] \, x^{\Delta_a-1} \big|_{x=1} \,.
\end{align}
The \emph{Zhu algebra} $\Zhu_H V$ of a vertex algebra $V$ with a 
Hamiltonian operator $H$ is defined as the quotient $V/J$, where 
\begin{equation*}
J=\Span_\CC\bigl\{a*_{-2}b\,|\,a,b\in V\bigr\} \,.
\end{equation*}
Then $\Zhu_H V$ is an associative algebra with a product induced by
the $*_{-1}$-product on $V$ (see \cite[Theorem 2.1.1]{Z}).
This means that in $\Zhu_H V$ we have
\begin{equation*}
\piz(a)\,\piz(b) = \piz(a*_{-1}b) 
\,, \qquad a,b \in V \,,
\end{equation*}
where $\piz$ denotes the natural quotient map $V\to V/J$.
Moreover, by \cite[Eq.\ (2.1.4)]{Z}, we have
\begin{equation*}
\piz(a)\,\piz(b) - \piz(b)\,\piz(a) = \piz([a_* b]) 
\,, \qquad a,b \in V \,.
\end{equation*}

Assume now that $V=V(R)$ is the universal enveloping vertex algebra 
of a non-linear Lie conformal algebra $R$, so that the 
Hamiltonian operator $H$ on $V$ agrees with the grading of $R$
by conformal weight, i.e., $Ha=\Delta_a \, a$ for $a\in R\subset V$.
Then the $*$-bracket \eqref{e-zhu1} is well defined on the quotient
$\mf z := R/(T+H)R$, and the induced operation $[\,\,,\,]$ on $\mf z$
endows it with the structure of
a \emph{non-linear Lie algebra} (see \cite{DSK2}).
Since $V$ is freely generated by $R$ and
\begin{equation*}
a*_{-2}\vac = a_{(-2)}\vac + \Delta_a \, a_{(-1)}\vac = (T+H)a 
\,, \qquad a\in V \,,
\end{equation*}
we can identify $\mf z$ as a subspace of $\Zhu_H V$.
Then by \cite[Corollary 3.26]{DSK2}, the Zhu algebra
$\Zhu_H V$ is an associative algebra generated by $\mf z$
with relations $ab-ba=[a,b]$ for $a,b\in\mf z$.

\subsection{The Zhu algebra of $V^d_{-1}$}
Recall that $V^d_{-1}=V(R_{-1}^d)$ is the universal enveloping vertex algebra
of the non-linear Lie conformal algebra $R_{-1}^d$ defined in 
Theorem \ref{th-june12}. Introduce a Hamiltonian operator $H$ on $V_{-1}^d$
such that $\Delta_h=1$ and $\Delta_e+\Delta_f=d+1$ (cf.\ \eqref{e1.2}).
Once we fix the value of $\Delta_e$, this determines $H$ uniquely,
because the vertex algebra $V_{-1}^d$ is strongly generated by 
the elements $h,e,f$.

For any polynomial $p(\la)\in\CC[\la]$, S.P.~Smith investigated in \cite{S}
the associative algebra with generators $h,e,f$ and relations
\begin{equation*}
he-eh = e \,, \qquad hf-fh = -f \,, \qquad ef-fe = p(h) \,.
\end{equation*}
We will prove that $\Zhu_H V_{-1}^d$ is one of Smith's algebras.

\begin{theorem}\label{t-zhu}
The Zhu algebra $\Zhu_H V_{-1}^d$ is the associative algebra with generators 
$h,e,f$ and relations
\begin{equation*}
[h,e] = e \,, \qquad [h,f] = -f \,, \qquad
[e,f] = (\Delta_e-h-1)\cdots(\Delta_e-h-d) \,,
\end{equation*}
where  $[a,b]=ab-ba$ denotes commutator with respect to the 
associative product.
\end{theorem}

In order to prove Theorem \ref{t-zhu}, we utilize the results of \cite{DSK2}
discussed at the end of the previous subsection. Since
$R_{-1}^d=\CC[T]\langle h,e,f \rangle$, the space $\mf z=R/(T+H)R$
is three-dimensional with a basis $\{h,e,f\}$. Here and further, we
identify the elements $h,e,f \in R_{-1}^d \subset V_{-1}^d$ with their
images under the projection $\piz\colon V\to \Zhu_H V_{-1}^d$.
By \cite[Corollary 3.26]{DSK2}, $\Zhu_H V_{-1}^d$ is the 
associative algebra generated by $h,e,f$ subject to the
relations $ab-ba=[a,b]=\piz([a_*b])$.
It follows immediately from the definition \eqref{e-zhu1} that
$[h_*e]=e$ and $[h_*f]=-f$.
Therefore, to prove Theorem \ref{t-zhu}, we are left to check that
\begin{equation}\label{e-zhu2}
\piz([e_*f]) = d! \, \binom{\Delta_e-h-1}{d}\,.
\end{equation}
We are going to use the following lemma.

\begin{lemma}\label{l-zhu1}
For every $s\geq1$ and $n_1,\dots,n_s\in\ZZ_+$, we have 
\begin{equation*}
\piz\bigl({:}(T^{n_1} h)\cdots(T^{n_s}h){:}\bigr)
= \piz(T^{n_1} h) \cdots \piz(T^{n_s} h)
=(-1)^{n_1+\cdots+n_s} \, n_1!\cdots n_s! \, h^s\,.
\end{equation*}
\end{lemma}
\begin{proof}
Notice that equations \eqref{e-zhu1a} and \eqref{vac2} imply 
$a*_{-2}\vac = Ta+\Delta_a \, a$, and hence $\piz(Ta)=-\Delta_a \, \piz(a)$. 
Using this and $\Delta_{Ta}=\Delta_a+1$, 
it is easy to check by induction that 
$\piz(T^nh)=(-1)^n \, n! \, h$, which is exactly the statement of the lemma
for $s=1$. We will prove the general case by induction on $s$. 
Letting $A={:}(T^{n_2} h)\cdots(T^{n_s}h){:}$, we compute
\begin{align*}
(T^n h)*_{-1}A &= \sum_{j\in\ZZ_+}\binom{\Delta_{T^nh}}{j} \, (T^nh)_{(j-1)}A 
\\
& ={:}(T^n h)A{:}+\sum_{k=0}^n\binom{n+1}{k+1} \, (T^nh)_{(k)}A 
\\
&={:}(T^n h)A{:}+(-1)^n \, n! \, h_{(0)}A \,,
\end{align*}
using that by sesqui-linearity $(Ta)_{(k)} b = -k \, a_{(k-1)} b$.
However, since $h_{(0)}h=0$, we have $h_{(0)}A=0$. Thus,
\begin{equation*}
\piz({:}(T^n h)A{:})
=\piz((T^n h)*_{-1}A)=\piz(T^n h) \, \piz(A)=(-1)^n \, n! \, h \, \piz(A)\,,
\end{equation*}
completing the proof of the lemma.
\end{proof}
\begin{proof}[Proof of Theorem \ref{t-zhu}]
As explained above, we only need to check equation \eqref{e-zhu2}.
By the definition \eqref{e-zhu1} of the $*$-bracket, we have
\begin{equation*}
\piz([e_* f])
=\piz\bigl([e_{\partial_x} f] \, x^{\Delta_e-1} \big|_{x=1} \bigr) 
=\piz([e_{\partial_x} f]) \, x^{\Delta_e-1} \big|_{x=1}\,,
\end{equation*}
where $[e_{\partial_x} f]$ is obtained by replacing $\la$ with $\partial_x$
in the $\la$-bracket $[e_\la f]$.
Recall that, by \eqref{nlie5} and \eqref{e3.13} for $\beta=-1$, 
$\gamma=d!$, we have
\begin{equation*}
[e_\la f] = {:}(\lambda+T-h)^d \, 1{:} 
= d! \, \Res_z \frac{e^{z\lambda}}{z^{d+1}} \,
{:} \exp \Bigl( -\sum_{k=1}^\infty \frac{z^k}{k!} \, T^{k-1}h \Bigr) {:} \,.
\end{equation*}
Applying Lemma \ref{l-zhu1}, we obtain
\begin{align*}
\frac1{d!} \, \piz([e_\la f]) 
&= \Res_z \frac{e^{z\lambda}}{z^{d+1}} \,
\exp \Bigl( -\sum_{k=1}^\infty \frac{z^k}{k!} \, \piz(T^{k-1}h) \Bigr) 
\\
&= \Res_z \frac{e^{z\lambda}}{z^{d+1}} \,
\exp \Bigl( \sum_{k=1}^\infty \frac{(-z)^k}{k} \, h \Bigr)
= \Res_z \frac{e^{z\lambda}}{z^{d+1}} \, (1+z)^{-h} \,,
\end{align*}
using the Taylor expansion of $\log(1+z)$.
Then
\begin{align*}
\frac1{d!} \, \piz([e_* f]) 
&= \Res_z \frac{e^{z\partial_x}}{z^{d+1}} \,
(1+z)^{-h} \, x^{\Delta_e-1}\big|_{x=1}
\\
&= \Res_z z^{-d-1} \, (1+z)^{-h+\Delta_e-1} 
= \binom{-h+\Delta_e-1}{d} \,,
\end{align*}
which completes the proof of the theorem.
\end{proof}

\subsection{The Zhu algebra of $V_{\ZZ\sqrt\be}$}
The computations of the previous subsection can also be used to determine
the Zhu algebra of the lattice vertex algebra $V_{\ZZ\sqrt\be}$,
where $\be=d+1 \in 2\NN$ (see Section \ref{svbe}).
Here we assume $\be$ is even, so that the generators $e$ and $f$ are even, and
\begin{equation*}
\Delta_h=1 \,, \qquad \Delta_e=\Delta_f=\be/2 \,.
\end{equation*}
By Remark \ref{r-july17}, the $\la$-bracket $[e_\la f]$ in $V_{\ZZ\sqrt\be}$
is obtained by taking $\be=d+1$, $\gamma=1$ in formula \eqref{e3.13}.
Then the above proof of Theorem \ref{t-zhu}
gives that $\Zhu_H V_{\ZZ\sqrt\be}$ is a quotient
of the associative algebra with generators $h,e,f$ and relations
\begin{equation*}
[h,e] = e \,, \qquad [h,f] = -f \,, \qquad
[e,f] = \binom{\be/2+\be h-1}{\be-1} \,.
\end{equation*}
To obtain $\Zhu_H V_{\ZZ\sqrt\be}$, we need
to quotient by the additional relations \eqref{vbe5}, i.e.,
by the elements $\piz(Te-\be\, {:}he{:})$ and $\piz(Tf+\be\, {:}hf{:})$.
As in the proof of Lemma \ref{l-zhu1}, we have: 
$\piz(Te) = -\Delta_e \, e = -\be e/2$ and $\piz(Tf)=-\be f/2$.
On the other hand, by \eqref{e-zhu1a} and \eqref{e1.1},
\begin{equation*}
h*_{-1}e = h_{(-1)}e + h_{(0)}e = {:}he{:} + e \,, \qquad
h*_{-1}f = {:}hf{:} - f \,.
\end{equation*}
Hence, in $\Zhu_H V_{\ZZ\sqrt\be}$ we have the relation
\begin{equation*}
he = \piz(h*_{-1}e) = \piz({:}he{:}) + e 
= \frac1\be \, \piz(Te) + e = \frac{e}2 \,,
\end{equation*}
and similarly $hf=-f/2$.
Letting $\be=2k$, we obtain precisely the result of \cite[Theorem 3.2]{DLM1}, 
since
\begin{equation*}
\binom{k+2kh-1}{2k-1}
= \frac{2k}{(2k-1)!} \, h(4k^2h^2-1)(4k^2h^2-4)\cdots(4k^2h^2-(k-1)^2) \,.
\end{equation*}

\begin{remark}\label{rem-zhu2}
The above result of \cite{DLM1} shows that the Zhu algebra of the lattice 
vertex algebra $V_{\ZZ\sqrt\be}$ is a quotient of a Smith algebra.
In \cite[Remark 3.3]{DLM1}, the authors asked whether one can find a vertex 
algebra whose Zhu algebra is a Smith algebra and which has $V_{\ZZ\sqrt\be}$ 
as a quotient. By Remark \ref{rem-june12}, the answer to that question is 
negative.
\end{remark}


\section*{Acknowledgments}
We are grateful to Victor Kac for teaching us many things, 
among which vertex algebras and conformal algebras.
We thank the referee for several suggestions for improving the exposition, 
and M.S.~Plyushchay for pointing out \cite{AP,LP} where non-linear
extensions of superconformal quantum mechanics were considered.


\end{document}